\documentclass[11pt]{article}

\usepackage[english]{babel}
\usepackage[latin1]{inputenc}
\usepackage{lmodern,csquotes}
\usepackage[T1]{fontenc}

\usepackage{amssymb,amsmath,amsthm,latexsym,amsfonts,amscd,dsfont,enumerate,bbm}
\usepackage{a4wide,graphicx,tikz}

\usepackage[tablesfirst,nomarkers,nolists]{endfloat}
\renewcommand{\efloatheading}[1]{}

\usepackage[backend=biber,style=authoryear-comp,maxcitenames=1,sorting=nyt,sortcites=false]{biblatex}
\usepackage{pgfplots,pgfplotstable,booktabs,colortbl}
\pgfplotsset{compat=newest}
\usepgfplotslibrary{fillbetween}

\pgfplotsset{
  layers/axis lines on top/.define layer set={
    axis background,
    axis grid,
    axis ticks,
    axis tick labels,
    pre main,
    main,
    axis lines,
    axis descriptions,
    axis foreground,
  }{/pgfplots/layers/standard},
}

\pgfplotstableread{data/TTF_200114_band.dat}\ttfaband

\pgfplotstableread{data/Cocoa_211215_market.dat}\cocoamkt
\pgfplotstableread{data/Coffee_211215_market.dat}\coffeemkt
\pgfplotstableread{data/Sugar_211215_market.dat}\sugarmkt
\pgfplotstableread{data/Gold_211215_market.dat}\goldmkt
\pgfplotstableread{data/Silver_211215_market.dat}\silvermkt
\pgfplotstableread{data/Copper_211215_market.dat}\coppermkt
\pgfplotstableread{data/Soybean_211215_market.dat}\soybeanmkt
\pgfplotstableread{data/Wheat_211215_market.dat}\wheatmkt
\pgfplotstableread{data/HeatingOil_211215_market.dat}\heatingoilamkt
\pgfplotstableread{data/HeatingOil_220311_market.dat}\heatingoilbmkt
\pgfplotstableread{data/Brent_211215_market.dat}\brentmkt
\pgfplotstableread{data/WTI_211215_market.dat}\wtimkt
\pgfplotstableread{data/NBP_220311_market.dat}\nbpmkt
\pgfplotstableread{data/NG_220311_market.dat}\ngmkt
\pgfplotstableread{data/TTF_200114_market.dat}\ttfamkt
\pgfplotstableread{data/TTF_201228_market.dat}\ttfbmkt
\pgfplotstableread{data/TTF_211215_market.dat}\ttfcmkt
\pgfplotstableread{data/TTF_220311_market.dat}\ttfdmkt
\pgfplotstableread{data/GermanyPower_211215_market.dat}\germanypowermkt

\pgfplotstableread{data/Cocoa_211215_mesh.dat}\cocoamsh
\pgfplotstableread{data/Coffee_211215_mesh.dat}\coffeemsh
\pgfplotstableread{data/Sugar_211215_mesh.dat}\sugarmsh
\pgfplotstableread{data/Gold_211215_mesh.dat}\goldmsh
\pgfplotstableread{data/Silver_211215_mesh.dat}\silvermsh
\pgfplotstableread{data/Copper_211215_mesh.dat}\coppermsh
\pgfplotstableread{data/Soybean_211215_mesh.dat}\soybeanmsh
\pgfplotstableread{data/Wheat_211215_mesh.dat}\wheatmsh
\pgfplotstableread{data/HeatingOil_211215_mesh.dat}\heatingoilamsh
\pgfplotstableread{data/HeatingOil_220311_mesh.dat}\heatingoilbmsh
\pgfplotstableread{data/Brent_211215_mesh.dat}\brentmsh
\pgfplotstableread{data/WTI_211215_mesh.dat}\wtimsh
\pgfplotstableread{data/NBP_220311_mesh.dat}\nbpmsh
\pgfplotstableread{data/NG_220311_mesh.dat}\ngmsh
\pgfplotstableread{data/TTF_200114_mesh.dat}\ttfamsh
\pgfplotstableread{data/TTF_201228_mesh.dat}\ttfbmsh
\pgfplotstableread{data/TTF_211215_mesh.dat}\ttfcmsh
\pgfplotstableread{data/TTF_220311_mesh.dat}\ttfdmsh
\pgfplotstableread{data/GermanyPower_211215_mesh.dat}\germanypowermsh

\definecolor{mylightyellow}{rgb}{1,1,.8}
\definecolor{mylightgreen}{rgb}{.8,1,.8}
\definecolor{mydarkred}{RGB}{178,34,34}
\definecolor{mydarkgreen}{RGB}{34,139,34}
\definecolor{mydarkblue}{RGB}{72,61,139}
\definecolor{mydarkyellow}{RGB}{218,165,32}
\definecolor{mydarkmagenta}{RGB}{231,84,128}

\definecolor{myblueA}{RGB}{52,41,39}
\definecolor{myblueB}{RGB}{92,81,109}
\definecolor{myblueC}{RGB}{132,121,179}
\definecolor{myblueD}{RGB}{172,161,249}

\usepackage[]{hyperref}
 \hypersetup{
 colorlinks=true,breaklinks=true,
 urlcolor= mydarkblue,linkcolor= mydarkblue,citecolor= mydarkgreen,
 pdfauthor={Maran, A., Pallavicini, A.},
}

\newcommand{\Eq}[1]{{\small\[{#1}\]}}

\newcommand{\Exo}[1]{\mathbb{E}\!\left[\,#1\,\right]}
\newcommand{\Ex}[2]{\mathbb{E}_{#1}\!\left[\,#2\,\right]}
\newcommand{\ExG}[2]{\mathbb{E}\!\left[\,#2\,\left|{\cal G}_{#1}\right.\right]}
\newcommand{\ExF}[2]{\mathbb{E}\!\left[\,#2\,\left|{\cal F}_{#1}\right.\right]}
\newcommand{\ExFT}[3]{\mathbb{E}^{#2}\!\left[\,#3\,\left.\right|{\cal F}_{#1}\right]}
\newcommand{\ExGT}[3]{\mathbb{E}^{#2}\!\left[\,#3\,\left.\right|{\cal G}_{#1}\right]}
\newcommand{\ExCo}[2]{\mathbb{E}\!\left[\left.\,#1\,\right|\,#2\,\right]}
\newcommand{\ExC}[3]{\mathbb{E}_{#1}\!\left[\left.\,#2\,\right|\,#3\,\right]}
\newcommand{\ExCT}[4]{\mathbb{E}_{#1}^{#2}\!\left[\left.\,#3\,\right|\,#4\,\right]}
\newcommand{\ExT}[3]{\mathbb{E}_{#1}^{#2}\!\left[\,#3\,\right]}
\newcommand{\Exf}[2]{\hat{\mathbb{E}}_{#1}\!\left[\,#2\,\right]}
\newcommand{\ExfT}[3]{\hat{\mathbb{E}}_{#1}^{#2}\!\left[\,#3\,\right]}
\newcommand{\Px}[1]{\mathbb{P}\left\{\,#1\,\right\}}
\newcommand{\PxC}[2]{\mathbb{P}\!\left\{\,#1\,\left.\right|\,#2\,\right\}}
\newcommand{\Qx}[1]{\mathbb{Q}\left\{\,#1\,\right\}}
\newcommand{\QxC}[2]{\mathbb{Q}\!\left\{\,#1\,\left.\right|\,#2\,\right\}}
\newcommand{\QxCT}[3]{\mathbb{Q}^{#1}\!\left\{\,#2\,\left.\right|\,#3\,\right\}}
\newcommand{\QxT}[2]{\mathbb{Q}^{#1}\left\{\,#2\,\right\}}
\newcommand{\Qxf}[1]{\hat{\mathbb{Q}}\left\{\,#1\,\right\}}
\newcommand{\ind}[1]{1_{\{#1\}}}
\newcommand{\one}{\ensuremath{\mathbbm{1}}}
\newcommand{\onehalf}{\tfrac{1}{2}}
\newcommand{\var}[3]{{\rm Var}_{#1}^{#2}\!\left[\,#3\,\right]}
\newcommand{\cov}[4]{{\rm Cov}_{#1}^{#2}\!\left[\,#3\,,\,#4\,\right]}
\newcommand{\corr}[4]{{\rm Corr}_{#1}^{#2}\!\left[\,#3\,,\,#4\,\right]}

\newcommand{\rec}{R}
\newcommand{\lgd}{\mbox{L{\tiny GD}}}

\newcommand{\CCorr}[2]{{\rm Corr}\!\left[\,#1,#2\,\right]}

\newcommand{\pvcall}{{\mathcal{C}}}
\newcommand{\pvput}{{\mathcal{P}}}

\newcommand{\argmin}[1]{\underset{{#1}}{\operatorname{arg\,min}}\,}
\newcommand{\argmax}[1]{\underset{{#1}}{\operatorname{arg\,max}}\,}
\newcommand{\trig}[1]{\operatorname{trig}_{#1}}
\newcommand{\transpose}[1]{{#1}^{\intercal}} 
\newcommand{\clip}[3]{\operatorname{clip}\!\left({#1},{#2},{#3}\right)}

\title{Interpolating commodity futures prices with Kriging}
\author{
Andrea Maran\thanks{Independent consultant}
\and
Andrea Pallavicini\thanks{Intesa SanPaolo Milan, {\tt andrea.pallavicini@intesasanpaolo.com}.}
}
\date{
\small First Version: December 16, 2020.  This version: \today
}

\begin{document}

\maketitle

\begin{abstract}

The shape of the futures term structure is essential to commodity hedgers and speculators as futures prices serve as a forecast of future spot prices. Commodity markets quotes futures prices on a selection of maturities and delivery periods. In this note, we investigate a Bayesian technique known as Kriging to build a term structure of futures prices by embedding trends and seasonalities and by taking into account bid-ask spreads of market quotations on different delivery periods.

\end{abstract}

\bigskip

\noindent {\bf JEL classification codes:} C11, G13.\\
\noindent {\bf AMS classification codes:} 62F15, 91G20, 91G70.\\
\noindent {\bf Keywords:} Kriging, Commodity Futures, Futures Term Structure, Seasonality patterns.

\newpage
{\small \tableofcontents}
\vfill
{\footnotesize \noindent The opinions here expressed  are solely those of the authors and do not represent in any way those of their employers.}
\newpage

\maketitle

\pagestyle{myheadings} \markboth{}{{\footnotesize A. Maran, A. Pallavicini, Interpolating commodity futures prices with Kriging}}

\section{Introduction}

Commodity markets quote the prices of futures contracts for different maturity dates and in some cases for different delivery periods. For instance, futures contracts on TTF natural gas have a delivery period ranging from one month up to one year. Market quotes display complex patterns due also to the physical nature of the underlying assets. These patterns may feature periodic structures, usually refererred to as seasonsal effects. We refer to \textcite{Geman2009} for an overview. The resulting shape of the term structure of futures prices may prevent the use of simple interpolating techniques usually adopted in other asset classes to achieve robust results. For instance, the impact of seasonal patterns in commodity prices is usually addressed by introducing deterministic trigonometric functions with different time scales. For example, we refer to the work of \textcite{Lucia2002}, where the analysis is performed in the Scandinavian electricity market, and the work of \textcite{Geman2005} focusing on soybean. In \textcite{Cartea2005,Cartea2012} the trigonometric function is replaced by a fifth-order Fourier series and a long-term trend is added. More details can be found in \textcite{Moreno2019}.

In this paper, we take a different point of view and we investigate a Bayesian technique known as Kriging originally developed for geophysical problems by \textcite{Krige1951,Matheron1963}. In a pioneering work, \textcite{Benth2015} used classical Kriging to interpolate energy futures on a single delivery period. Here, we extend this approach by using a more general form of Kriging, which allows us to consider at the same time market quotes of futures contracts with different delivery periods and to take into account the bid-ask spreads of market quotes. This approach is inspired by recent works of \textcite{Cousin2016,Maatouk2017} on bootstrapping volatility surfaces. We apply our framework to different commodity markets including the challenging setting of natural gas futures where seasonalities and heterogeneous contracts are key ingredients in the term-structure construction. For this goal, we develop in Section~\ref{sec:kriging} an original non-parametric algorithm to preserve the observed shape of the term structure over time, and in in Section~\ref{sec:numerics} we present a numerical investigation in different commodity markets.

\section{Interpolating noisy data with Kriging}
\label{sec:kriging}

Commodity markets usually quotes futures contracts prices. We can find quotes for futures contracts on different delivery periods, and also spread contracts between futures with different maturities. For instance, TTF natural gas is quoted with delivery periods of one, three, six and twelve months, along with spread between consecutive one-month or three-month contracts. In this case we can see that, if we introduce a model for the shortest delivery period (one month), we can model also the longer periods by averaging the one-month prices. In general we can assume to model the shortest delivery period which allows us to recover all the market quotes. We define the term structure of futures prices with shortest delivery period as the curve
\begin{equation}
  T \mapsto F(T)
\end{equation}%

The specific form of the term structure $F(T)$ depends on our modelling assumptions, and it must be fitted to market prices which are quoted only at a discrete set of maturity dates and in term of bid and ask prices. The shape of the term structure can be complex since it is sensitive not only to the activity of hedgers and speculators, but also it reflects the physical properties of the commodity, and the level of supply and demand in the markets. The resulting shapes may display periodic patterns over specific time scales. For instance, there is more natural gas demand (for heating) in winter than in summer. We term in the following such patterns as seasonalities. In order to correctly describe the seasonalities we need market quotes with maturities on a time grid finer than the typical scale of the seasonality itself. For example, natural gas prices display a regular pattern over one year, which we can easily read from the market for the first year since monthly quotes are present. Yet, futures price for contracts with longer maturities are usually not quoted, or if present quoted on wider delivery periods. Thus, the term structure model must face the problem to extrapolate the seasonal pattern to longer maturities.

Thus, if we wish to build a complete term structure for futures prices, we need (i) to model different complex shapes, (ii) to deal with seasonalities, (iii) to consider bid-ask spreads, (iv) to bootstrap quotes from the prices of overlapping linear products. In the following, we consider Kriging, a Bayesian approach, to include all these features.

\subsection{Description of the model}

The basic idea of Kriging is to predict the value of an unknown function at a given point by computing a weighted average of the known values of the function in the neighbourhood of the point. Originally developed for geophysical problems by \textcite{Krige1951,Matheron1963}, this technique is also known as Gaussian process regression or Wiener-Kolmogorov prediction. We wish to use Kriging as an effective interpolating technique for the futures term structure. Here, we extend the techniques developed in \textcite{Cousin2016,Maatouk2017} for zero-rate curves and volatility term structures to deal with our case of commodity futures prices.

We start by describing the model for the prices of futures contracts along with their no-arbitrage constraints. On the market we can find ask quotes $q^a \in \{ q^a_1,\ldots,q^a_n \}$ and bid quotes $q^b \in \{ q^b_1,\ldots,q^b_n \}$ of futures contracts on different delivery periods (month, quarter, season, year), and also of contracts on their spreads. Here, we consider only quotes on delivery periods of one or multiple months for ease of discussion. The extension to shorter tenors is straightforward. We introduce a term structure of future prices $F(T)$ for contracts with one-month delivery period and maturity $T \in \{ T_1,\ldots,T_m \}$. The quotes observed in the market shall satisfy the following no-arbitrage relationships
\begin{equation}
q^b_j \le \sum_{i=1}^m A_{ij} F(T_i) \le q^a_j
\end{equation}%
where the coefficients $A_{ij}$ depend on the contract. For instance, if the $j$-th futures contract has a delivery period of three months starting in $T_3$ and terminating in $T_6$, the coefficients are $A_{ij}=(T_i-T_{i-1})/(T_6-T_4)$ for $i\in\{4,5,6\}$ and zero for the other values of $i$. We call $\cal C$ the set of all the no-arbitrage constraints of the previous form.

For each bid-ask pair we can define the mid price $q_j := (q^a_j + q^b_j)/2$. The price we are modelling lies between the bid and ask prices, due to the previous constraint, but it may be different from the mid price. We can express this uncertainty by introducing a normal random variable $\epsilon_j$ for each $j\in\{1,\ldots,n\}$ such that
\begin{equation}
\label{eq:model}
  \sum_{i=1}^m A_{ij} F(T_i) - q_j = \epsilon_j
  \;,\quad
  \epsilon := (\epsilon_1,\ldots,\epsilon_n) \sim {\cal N}(0,\Sigma)
\end{equation}%
where the $n\times n$ covariance matrix $\Sigma$ can be defined from the bid-ask spreads of the quotes. In the numerical section we use the form $\Sigma_{j_1j_2} := \delta_{j_1j_2} (q^a_{j_1}-q^b_{j_1})^2/4$.

We choose to model the term structure $F(T)$ by means of a linear B-spline on an evenly spaced grid $\{t_1,\ldots,t_N\}$ with Gaussian coefficients independent of the price uncertainties, so that we get for $i\in\{1,\ldots,m\}$
\begin{equation}
\label{eq:term}
  F(T_i) := \sum_{k=1}^N \xi_k \,\phi(T_i-t_k)
  \;,\quad
  \xi := (\xi_1,\ldots,\xi_N) \sim {\cal N}(0,\Gamma)
\end{equation}%
with the function $\phi(x)$ defined as
\begin{equation}
  \phi(x) := \left( 1-\frac{|x|}{\delta t} \right)^+
\end{equation}%
where for any $k\in\{1,\ldots,N\}$ we can define $\delta t := t_{k+1} - t_k$, since the time-grid is evenly spaced, and the $N\times N$ covariance matrix $\Gamma$ is defined in term of a radial kernel $K$, and its coefficients $\Gamma_{k_1k_2}$ with $k_1,k_2 \in \{1,\ldots,N\}$ are given by
\begin{equation}
  \Gamma_{k_1k_2} := \sigma^2 K(|t_{k_1}-t_{k_2}|,\theta)
  \;,\quad
  K(x,\theta) := e^{-\frac{x^2}{2\theta^2}}
\end{equation}
with hyper-parameters $\sigma>0$ and $\theta>0$. The specific choice of using a B-splines to interpolate futures prices is due to the fact that it allows to deal in a simple way with the no-arbitrage constraints. Indeed, in this way the constraints remain linear in $\xi$. On the other hand, the choice of the radial kernel is a parsimonious parametrization usually employed in the literature. We tried different forms without any relevant advantage.

\subsection{Estimation of model parameters}

Now, we can proceed by illustrating the procedure we adopt to estimate the model parameters $\{\xi_1,\ldots,\xi_N\}$. We calculate the mode of the posterior distribution $\PxC{\xi}{q}$ of model parameters given the observations under the no-arbitrage constraints.
\begin{eqnarray}
\label{eq:estimate}
  {\bar\xi}
  &:=& \argmax{\xi\in{\cal C}} \PxC{\xi}{q} \\\nonumber 
  & =& \argmin{\xi\in{\cal C}} \left( \,\sum_{k_1,k_2=1}^N \xi_{k_1} ({\bar\Gamma}^{-1})_{k_1k_2} \xi_{k_2} + \sum_{j_1,j_2=1}^n \epsilon_{j_1} (\Sigma^{-1})_{j_1j_2} \epsilon_{j_2} \right)
\end{eqnarray}%
where $\epsilon_j$ can be calculated in term of the market mid price $q_j$ by using equation \eqref{eq:model}, and $\bar\Gamma$ is the matrix $\Gamma$ calculated starting from the optimal hyper-parameters $ \{{\bar\sigma},{\bar\theta}\}$ obtained by maximizing the unconstrained observation likelihood.
\begin{eqnarray}
  \{{\bar\sigma},{\bar\theta}\}
  &:=& \argmax{\sigma,\theta} \Px{q} \\\nonumber
  & =& \argmin{\sigma,\theta} \left( \log \det{C} + \sum_{j_1j_2} q_{j_1} (C^{-1})_{j_1j_2} q_{j_2} \right)
\end{eqnarray}%
where the $n\times n$ covariance matrix $C$ is given by
\begin{equation}
  C_{j_1j_2} := \Sigma_{j_1j_2} + \sum_{i_1,i_2=1}^m A_{i_1j_1} A_{i_2j_2} \sum_{k_1,k_2=1}^N \phi(T_{i_1}-t_{k_1}) \,\Gamma_{k_1k_2} \phi(T_{i_2}-t_{k_2})
\end{equation}%
Once the estimation procedure is completed we can use the optimal parameters $\{\bar\xi_1,\ldots,\bar\xi_N\}$ to calculate the futures prices by using equation \eqref{eq:term}.

\subsection{Treatment of seasonalities}

Futures prices quoted in commodity markets may display regular variations over the year. We focus here for ease of discussion on the case of the TTF natural gas. In this case the market quotes the prices of futures contracts with a delivery period of one month and maturities on each month of the current year. A direct inspection of these quotes shows a regular pattern with cost of gas increasing in colder months. Yet, for maturities longer than one year we do not find quotes with a delivery on a single month, but on three months or even on longer periods. Thus, we should face the problem to replicate the pattern found in the first year on the following ones.

We can include seasonalities in Kriging by penalizing the minimization used to estimate the optimal parameters so that the shape of the term structure is preserved. We modify equation \eqref{eq:estimate} in the following way
\begin{equation}
\label{eq:estimate-seasonality}
  {\bar\xi} := \argmin{\xi\in{\cal C}} \left( \ldots + \gamma \sum_{k:\,t_k>{\rm 1y}} \left( \partial_T^2 F(T)|_{T=t_k} - \partial_T^2 F(T)|_{T=t_{m(k)}} \right)^2 \right)
\end{equation}%
where $\gamma$ is a penalty parameter, and $m(k)$ returns an index such that $t_{m(k)}$ occurs in the first year on the same month and day of $t_k$. Thanks to the choice of interpolating the futures prices by means of B-splines, we can explicitly calculate the above derivatives as
\begin{equation}
  \partial_T^2 F(T)|_{T=t_k} \approx \frac{1}{\delta t^2} \left( F(t_{k-1}) - 2 F(t_k) + F(t_{k+1}) \right) = \frac{1}{\delta t^2} \left( \xi_{k-1} - 2 \xi_k + \xi_{k+1} \right)
\end{equation}%

\section{Numerical investigations}
\label{sec:numerics}

We illustrate the model capabilities by investigating some commodity markets. The performance of the model ranges from one tenth of second to two seconds per calibration depending on the number of market quotes. For each data-set we calibrate our model with and without the seasonality management. We name in figures and tables the model prices with $F$ if seasonality is managed (continuous blue lines in figures), otherwise we name the prices $F^{\rm K}$ (dotted blue lines in figures). The optimization problems used for the estimations are respectively equations \eqref{eq:estimate-seasonality} and \eqref{eq:estimate}. Moreover, we add as a comparison the results obtained with the model implemented in \textcite{Benth2015}, where futures prices are interpolated by means of classical Kriging. This approach has several limitations that our contribution wishes to overcome, but it is an interesting and relevant benchmark. We recall that this approach models only mid prices of futures contracts with a specific delivery period, for instance one month. We name this benchmark prices $F^{\rm B}$ (continuous pink lines in figures). In the tables we list bid and ask prices quoted in the market, model implied price with and without seasonality management, and benchmark prices implied by classical Kriging. In this way is it possible to check if the models fulfill arbitrage requirements.

In all the cases we notice that our model is able to keep a stable pattern even in presence of large bid-ask spreads and at longer maturities when fewer quotes are listed. As expected, seasonal effects are better reproduced when the penalty term introduced in equation \eqref{eq:estimate-seasonality} is considered. On the other hand, classical Kriging of \textcite{Benth2015} seems working only when monthly quotes are listed seamless with tight bid-ask spreads. We details our results in the following sections.

\subsection{Natural gas}

We investigate natural gas prices on different observation dates and different trading points. In particular, we consider TTF from EEX on 14 January 2020 (table \ref{tab:ttfa}) and 28 December 2020 (table \ref{tab:ttfb}), TTF from ICE on 15 December 2021 (table \ref{tab:ttfc}) and 11 March 2022 (table \ref{tab:ttfd}), NBP from ICE on 11 March 2022 (table \ref{tab:nbp}), NG from NYMEX on 11 March 2022 (table \ref{tab:ng}).

We notice that the market can quote futures with different delivery periods (month, quarter, season, and year) and spreads between months or quarters. If we analyze data for TTF in 2020 we can see a yearly seasonal pattern superimposed on a small trend (figures \ref{fig:ttfa} and \ref{fig:ttfb}). Our model integrating the information from all the market delivery periods is able to reproduce the seasonal pattern of the first year in the following ones. On the contrary, classical Kriging, which can be calibrated only to one-month quotes, fails to show a reasonable behaviour for longer maturities. When we look at TTF data in late 2021 and in 2022, when the market register a big distress due to political turmoils, we can see that our model is able to follow both the sharp trend of market quotes for the first two years and the reappearing of a seasonal pattern for longer maturities (figures \ref{fig:ttfc} and \ref{fig:ttfd}). On the contrary, classical Kriging not only shows its limits for longer maturities, but it also displays a noisy pattern for shorter maturities. Results confirmed also when we look at gas traded at different trading points like NBP and NG (figures \ref{fig:nbp} and \ref{fig:ng}). We notice also that the penalty term introduced in equation \eqref{eq:estimate-seasonality} contributes to improve the model behaviour at longer maturities.

In the case of natural gas we further investigate market and model uncertainties. Our Bayesian approach allow us to obtain a posterior distribution for model parameters. In equations \eqref{eq:estimate} and \eqref{eq:estimate-seasonality} we calculate the mode of the distribution to fix the model parameters for the our best estimate of the futures term structure. On the other hand, we can also examine the distibution of the futures term structures generated by our model. In particular, we can sample from the posterior distribution to obatin an estimate of the term structure at a given quantile level. We show in figure \ref{fig:ttfa} the results for TTF from EEX on 14 January 2020 for the model with seasonal pattern management. The shaded areas around the futures term structure (continuous blue lines) are the 1-$\sigma$ confidence level obtained empirically by sampling from the unconstrained posterior distribution and rejecting results outside of no-arbitrage constraints. This area shows the sum of market and model uncertainties contributing to the model parameter posterior distribution. We notice that maturities without quoted futures show a bigger uncertainty, since possible term structures are not constrained by market data.

\begin{table}
\begin{center}
\pgfplotstabletypeset[
    col sep=tab,
    columns={id,bid,ask,Fplain,Fstag,Fbenth},
    columns/id/.style={column name=Contract, column type={c|}, string type},
    columns/bid/.style={column name=$q^b$, fixed, fixed zerofill, precision=2, dec sep align, clear infinite},
    columns/ask/.style={column name=$q^a$, fixed, fixed zerofill, precision=2, dec sep align, clear infinite},
    columns/Fplain/.style={column name=$F$, fixed, fixed zerofill, precision=2, dec sep align, clear infinite},
    columns/Fstag/.style={column name=$F^{\rm K}$, fixed, fixed zerofill, precision=2, dec sep align, clear infinite},
    columns/Fbenth/.style={column name=$F^{\rm B}$, fixed, fixed zerofill, precision=2, dec sep align, clear infinite},
    every row no 5/.style={before row=\midrule},
    every row no 9/.style={before row=\midrule},
    every row no 14/.style={before row=\midrule},
    every row no 17/.style={before row=\midrule},
    every row no 19/.style={before row=\midrule},
    every head row/.style={before row=\toprule,after row=\midrule},
    every last row/.style={after row=\bottomrule},
]{\ttfamkt}
\end{center}
\caption{Market data for TTF natural gas observed on EEX exchange on 14 January 2020. From top to bottom: one-month contracts, quarters (three-month contracts), seasons (summer is the average of the second and third quarter, winter is the average of the fourth quarter and the first one of the following year), years, spreads among months, spread among quarters. The columns report from left to right contract identifier, bid and ask prices, model implied prices with and without seasonality management, classical Kriging implied prices.}
\label{tab:ttfa}
\end{table}

\begin{figure}
\begin{center}
\scalebox{0.9}{%
\begin{tikzpicture}
\begin{axis}[set layers=axis lines on top,
             xlabel=Maturity,
             ylabel=Futures Price,
             ylabel style={overlay},
             xmin=-0.1, xmax=3.1,
             xtick={0,1.003,2.003,3.003},
             xticklabels={{\tt JAN20},{\tt JAN21},{\tt JAN22},{\tt JAN23}},
             ymin=7, ymax=23,
             grid=major,
             legend style={legend pos=south west},
             axis background/.style={fill=gray!10}]
\addplot [name path=lba,color=mydarkblue!25,smooth] table [y=1sigma_lb1,x=T] from \ttfaband;
\addplot [name path=uba,color=mydarkblue!25,smooth] table [y=1sigma_ub1,x=T] from \ttfaband;
\addplot [color=mydarkblue!25] fill between [of=lba and uba];
\addplot [color=mydarkblue,thick,smooth] table [y=Fstag1,x=T] from \ttfamsh;
\addplot [color=mydarkblue,thick,smooth,style=dotted] table [y=Fplain1,x=T] from \ttfamsh;
\addplot [color=mydarkmagenta,thick,smooth] table [y=Fbenth1,x=T] from \ttfamsh;
\addplot [color=mydarkred,thick,only marks,mark=triangle,mark options={yshift=(\pgfplotmarksize+\pgflinewidth)/2}] table [y=Mbid1,x=T] from \ttfamsh;
\addplot [color=mydarkred,thick,only marks,mark=triangle,mark options={yshift=(\pgflinewidth-\pgfplotmarksize)/2,style={rotate=180}}] table [y=Mask1,x=T] from \ttfamsh;
\end{axis}
\end{tikzpicture}}
\hspace*{0.5cm}
\scalebox{0.9}{%
\begin{tikzpicture}
\begin{axis}[set layers=axis lines on top,
             xlabel=Maturity,
             ylabel=Futures Price,
             ylabel style={overlay},
             xmin=-0.1, xmax=3.1,
             xtick={0,1.003,2.003,3.003},
             xticklabels={{\tt JAN20},{\tt JAN21},{\tt JAN22},{\tt JAN23}},
             ymin=7, ymax=23,
             grid=major,
             legend style={legend pos=south west},
             axis background/.style={fill=gray!10}]
\addplot [name path=lba,color=mydarkblue!25,smooth] table [y=1sigma_lb3,x=T] from \ttfaband;
\addplot [name path=uba,color=mydarkblue!25,smooth] table [y=1sigma_ub3,x=T] from \ttfaband;
\addplot [color=mydarkblue!25] fill between [of=lba and uba];
\addplot [color=mydarkblue,thick,smooth] table [y=Fstag3,x=T] from \ttfamsh;
\addplot [color=mydarkblue,thick,smooth,style=dotted] table [y=Fplain3,x=T] from \ttfamsh;
\addplot [color=mydarkmagenta,thick,smooth] table [y=Fbenth3,x=T] from \ttfamsh;
\addplot [color=mydarkred,thick,only marks,mark=triangle,mark options={yshift=(\pgfplotmarksize+\pgflinewidth)/2}] table [y=Mbid3,x=T] from \ttfamsh;
\addplot [color=mydarkred,thick,only marks,mark=triangle,mark options={yshift=(\pgflinewidth-\pgfplotmarksize)/2,style={rotate=180}}] table [y=Mask3,x=T] from \ttfamsh;
\end{axis}
\end{tikzpicture}}
\\\vspace*{0.5cm}
\scalebox{0.9}{%
\begin{tikzpicture}
\begin{axis}[set layers=axis lines on top,
             xlabel=Maturity,
             ylabel=Futures Price,
             ylabel style={overlay},
             xmin=-0.1, xmax=3.1,
             xtick={0,1.003,2.003,3.003},
             xticklabels={{\tt JAN20},{\tt JAN21},{\tt JAN22},{\tt JAN23}},
             ymin=7, ymax=23,
             grid=major,
             legend style={legend pos=south west},
             axis background/.style={fill=gray!10}]
\addplot [name path=lba,color=mydarkblue!25,smooth] table [y=1sigma_lb6,x=T] from \ttfaband;
\addplot [name path=uba,color=mydarkblue!25,smooth] table [y=1sigma_ub6,x=T] from \ttfaband;
\addplot [color=mydarkblue!25] fill between [of=lba and uba];
\addplot [color=mydarkblue,thick,smooth] table [y=Fstag6,x=T] from \ttfamsh;
\addplot [color=mydarkblue,thick,smooth,style=dotted] table [y=Fplain6,x=T] from \ttfamsh;
\addplot [color=mydarkmagenta,thick,smooth] table [y=Fbenth6,x=T] from \ttfamsh;
\addplot [color=mydarkred,thick,only marks,mark=triangle,mark options={yshift=(\pgfplotmarksize+\pgflinewidth)/2}] table [y=Mbid6,x=T] from \ttfamsh;
\addplot [color=mydarkred,thick,only marks,mark=triangle,mark options={yshift=(\pgflinewidth-\pgfplotmarksize)/2,style={rotate=180}}] table [y=Mask6,x=T] from \ttfamsh;
\end{axis}
\end{tikzpicture}}
\hspace*{0.5cm}
\scalebox{0.9}{%
\begin{tikzpicture}
\begin{axis}[set layers=axis lines on top,
             grid style={on layer=axis background},
             xlabel=Maturity,
             ylabel=Futures Price,
             ylabel style={overlay},
             xmin=-0.1, xmax=3.1,
             xtick={0,1.003,2.003,3.003},
             xticklabels={{\tt JAN20},{\tt JAN21},{\tt JAN22},{\tt JAN23}},
             ymin=7, ymax=23,
             grid=major,
             legend style={legend pos=south west},
             axis background/.style={fill=gray!10}]
\addplot [name path=lba,color=mydarkblue!25,smooth] table [y=1sigma_lb12,x=T] from \ttfaband;
\addplot [name path=uba,color=mydarkblue!25,smooth] table [y=1sigma_ub12,x=T] from \ttfaband;
\addplot [color=mydarkblue!25] fill between [of=lba and uba];
\addplot [color=mydarkblue,thick,smooth] table [y=Fstag12,x=T] from \ttfamsh;
\addplot [color=mydarkblue,thick,smooth,style=dotted] table [y=Fplain12,x=T] from \ttfamsh;
\addplot [color=mydarkmagenta,thick,smooth] table [y=Fbenth12,x=T] from \ttfamsh;
\addplot [color=mydarkred,thick,only marks,mark=triangle,mark options={yshift=(\pgfplotmarksize+\pgflinewidth)/2}] table [y=Mbid12,x=T] from \ttfamsh;
\addplot [color=mydarkred,thick,only marks,mark=triangle,mark options={yshift=(\pgflinewidth-\pgfplotmarksize)/2,style={rotate=180}}] table [y=Mask12,x=T] from \ttfamsh;
\end{axis}
\end{tikzpicture}}
\end{center}
\caption{Term structures of futures prices for TTF natural gas quoted on 14 January 2020 on EEX market. From top-left to bottom-right the panels show the delivery periods of one, three, six months and one year. Triangles represent bid-ask futures contracts prices. Continuous blue lines is our model with penalization, dotted blue lines without it, while continuous pink lines is classical Kriging. Shaded areas represent market and model uncertainties at 1-$\sigma$ confidence level around futures prices calculated with the model with penalization.}
\label{fig:ttfa}
\end{figure}

\begin{table}
\begin{center}
\pgfplotstabletypeset[
    col sep=tab,
    columns={id,bid,ask,Fplain,Fstag,Fbenth},
    columns/id/.style={column name=Contract, column type={c|}, string type},
    columns/bid/.style={column name=$q^b$, fixed, fixed zerofill, precision=2, dec sep align, clear infinite},
    columns/ask/.style={column name=$q^a$, fixed, fixed zerofill, precision=2, dec sep align, clear infinite},
    columns/Fplain/.style={column name=$F$, fixed, fixed zerofill, precision=2, dec sep align, clear infinite},
    columns/Fstag/.style={column name=$F^{\rm K}$, fixed, fixed zerofill, precision=2, dec sep align, clear infinite},
    columns/Fbenth/.style={column name=$F^{\rm B}$, fixed, fixed zerofill, precision=2, dec sep align, clear infinite},
    every row no 6/.style={before row=\midrule},
    every row no 11/.style={before row=\midrule},
    every row no 17/.style={before row=\midrule},
    every head row/.style={before row=\toprule,after row=\midrule},
    every last row/.style={after row=\bottomrule},
]{\ttfbmkt}
\end{center}
\caption{Market data for TTF natural gas observed on EEX exchange on 28 December 2020. From top to bottom: one-month contracts, quarters (three-month contracts), seasons (summer is the average of the second and third quarter, winter is the average of the fourth quarter and the first one of the following year), years. The columns report from left to right contract identifier, bid and ask prices, model implied prices with and without seasonality management, classical Kriging implied prices.}
\label{tab:ttfb}
\end{table}

\begin{figure}
\begin{center}
\scalebox{0.9}{%
\begin{tikzpicture}
\begin{axis}[xlabel=Maturity,
                    ylabel=Futures Price,
                    ylabel style={overlay},
                    xmin=-0.1, xmax=3.1,
                    xtick={0.085,1.085,2.085,3.085},
                    xticklabels={{\tt JAN21},{\tt JAN22},{\tt JAN23},{\tt JAN24}},
                    ymin=7, ymax=23,
                    grid=major,
                    legend style={legend pos=south west},
                    axis background/.style={fill=gray!10}]
\addplot [color=mydarkblue,thick,smooth] table [y=Fstag1,x=T] from \ttfbmsh;
\addplot [color=mydarkblue,thick,smooth,style=dotted] table [y=Fplain1,x=T] from \ttfbmsh;
\addplot [color=mydarkmagenta,thick,smooth] table [y=Fbenth1,x=T] from \ttfbmsh;
\addplot [color=mydarkred,thick,only marks,mark=triangle,mark options={yshift=(\pgfplotmarksize+\pgflinewidth)/2}] table [y=Mbid1,x=T] from \ttfbmsh;
\addplot [color=mydarkred,thick,only marks,mark=triangle,mark options={yshift=(\pgflinewidth-\pgfplotmarksize)/2,style={rotate=180}}] table [y=Mask1,x=T] from \ttfbmsh;
\end{axis}
\end{tikzpicture}}
\hspace*{0.5cm}
\scalebox{0.9}{%
\begin{tikzpicture}
\begin{axis}[xlabel=Maturity,
                    ylabel=Futures Price,
                    ylabel style={overlay},
                    xmin=-0.1, xmax=3.1,
                    xtick={0.085,1.085,2.085,3.085},
                    xticklabels={{\tt JAN21},{\tt JAN22},{\tt JAN23},{\tt JAN24}},
                    ymin=7, ymax=23,
                    grid=major,
                    legend style={legend pos=south west},
                    axis background/.style={fill=gray!10}]
\addplot [color=mydarkblue,thick,smooth] table [y=Fstag3,x=T] from \ttfbmsh;
\addplot [color=mydarkblue,thick,smooth,style=dotted] table [y=Fplain3,x=T] from \ttfbmsh;
\addplot [color=mydarkmagenta,thick,smooth] table [y=Fbenth3,x=T] from \ttfbmsh;
\addplot [color=mydarkred,thick,only marks,mark=triangle,mark options={yshift=(\pgfplotmarksize+\pgflinewidth)/2}] table [y=Mbid3,x=T] from \ttfbmsh;
\addplot [color=mydarkred,thick,only marks,mark=triangle,mark options={yshift=(\pgflinewidth-\pgfplotmarksize)/2,style={rotate=180}}] table [y=Mask3,x=T] from \ttfbmsh;
\end{axis}
\end{tikzpicture}}
\\\vspace*{0.5cm}
\scalebox{0.9}{%
\begin{tikzpicture}
\begin{axis}[xlabel=Maturity,
                    ylabel=Futures Price,
                    ylabel style={overlay},
                    xmin=-0.1, xmax=3.1,
                    xtick={0.085,1.085,2.085,3.085},
                    xticklabels={{\tt JAN21},{\tt JAN22},{\tt JAN23},{\tt JAN24}},
                    ymin=7, ymax=23,
                    grid=major,
                    legend style={legend pos=south west},
                    axis background/.style={fill=gray!10}]
\addplot [color=mydarkblue,thick,smooth] table [y=Fstag6,x=T] from \ttfbmsh;
\addplot [color=mydarkblue,thick,smooth,style=dotted] table [y=Fplain6,x=T] from \ttfbmsh;
\addplot [color=mydarkmagenta,thick,smooth] table [y=Fbenth6,x=T] from \ttfbmsh;
\addplot [color=mydarkred,thick,only marks,mark=triangle,mark options={yshift=(\pgfplotmarksize+\pgflinewidth)/2}] table [y=Mbid6,x=T] from \ttfbmsh;
\addplot [color=mydarkred,thick,only marks,mark=triangle,mark options={yshift=(\pgflinewidth-\pgfplotmarksize)/2,style={rotate=180}}] table [y=Mask6,x=T] from \ttfbmsh;
\end{axis}
\end{tikzpicture}}
\hspace*{0.5cm}
\scalebox{0.9}{%
\begin{tikzpicture}
\begin{axis}[xlabel=Maturity,
                    ylabel=Futures Price,
                    ylabel style={overlay},
                    xmin=-0.1, xmax=3.1,
                    xtick={0.085,1.085,2.085,3.085},
                    xticklabels={{\tt JAN21},{\tt JAN22},{\tt JAN23},{\tt JAN24}},
                    ymin=7, ymax=23,
                    grid=major,
                    legend style={legend pos=south west},
                    axis background/.style={fill=gray!10}]
\addplot [color=mydarkblue,thick,smooth] table [y=Fstag12,x=T] from \ttfbmsh;
\addplot [color=mydarkblue,thick,smooth,style=dotted] table [y=Fplain12,x=T] from \ttfbmsh;
\addplot [color=mydarkmagenta,thick,smooth] table [y=Fbenth12,x=T] from \ttfbmsh;
\addplot [color=mydarkred,thick,only marks,mark=triangle,mark options={yshift=(\pgfplotmarksize+\pgflinewidth)/2}] table [y=Mbid12,x=T] from \ttfbmsh;
\addplot [color=mydarkred,thick,only marks,mark=triangle,mark options={yshift=(\pgflinewidth-\pgfplotmarksize)/2,style={rotate=180}}] table [y=Mask12,x=T] from \ttfbmsh;
\end{axis}
\end{tikzpicture}}
\end{center}
\caption{Term structures of futures prices for TTF natural gas quoted on 28 December 2020 on EEX market. From top-left to bottom-right the panels show the delivery periods of one, three, six months and one year. Triangles represent bid-ask futures contracts prices. Continuous blue lines is our model with penalization, dotted blue lines without it, while continuous pink lines is classical Kriging.}
\label{fig:ttfb}
\end{figure}

\begin{table}
\begin{center}
\pgfplotstabletypeset[
    col sep=tab,
    columns={id,bid,ask,Fplain,Fstag,Fbenth},
    columns/id/.style={column name=Contract, column type={c|}, string type},
    columns/bid/.style={column name=$q^b$, fixed, fixed zerofill, precision=2, dec sep align, clear infinite},
    columns/ask/.style={column name=$q^a$, fixed, fixed zerofill, precision=2, dec sep align, clear infinite},
    columns/Fplain/.style={column name=$F$, fixed, fixed zerofill, precision=2, dec sep align, clear infinite},
    columns/Fstag/.style={column name=$F^{\rm K}$, fixed, fixed zerofill, precision=2, dec sep align, clear infinite},
    columns/Fbenth/.style={column name=$F^{\rm B}$, fixed, fixed zerofill, precision=2, dec sep align, clear infinite},
    every row no 12/.style={before row=\midrule},
    every row no 21/.style={before row=\midrule},
    every row no 27/.style={before row=\midrule},
    every head row/.style={before row=\toprule,after row=\midrule},
    every last row/.style={after row=\bottomrule},
]{\ttfcmkt}
\end{center}
\caption{Market data for TTF natural gas observed on ICE exchange on 15 December 2021. From top to bottom: one-month contracts, quarters (three-month contracts), seasons (summer is the average of the second and third quarter, winter is the average of the fourth quarter and the first one of the following year), years. The columns report from left to right contract identifier, bid and ask prices, model implied prices with and without seasonality management, classical Kriging implied prices.}
\label{tab:ttfc}
\end{table}

\begin{figure}
\begin{center}
\scalebox{0.9}{%
\begin{tikzpicture}
\begin{axis}[xlabel=Maturity,
                    ylabel=Futures Price,
                    ylabel style={overlay},
                    xmin=-0.1, xmax=5.1,
                    xtick={0.085,1.085,2.085,3.085,4.088,5.088},
                    xticklabels={{\tt JAN22},{\tt JAN23},{\tt JAN24},{\tt JAN25},{\tt JAN26},{\tt JAN27}},
                    ymin=7, ymax=133,
                    grid=major,
                    legend style={legend pos=south west},
                    axis background/.style={fill=gray!10}]
\addplot [color=mydarkblue,thick,smooth] table [y=Fstag1,x=T] from \ttfcmsh;
\addplot [color=mydarkblue,thick,smooth,style=dotted] table [y=Fplain1,x=T] from \ttfcmsh;
\addplot [color=mydarkmagenta,thick,smooth] table [y=Fbenth1,x=T] from \ttfcmsh;
\addplot [color=mydarkred,thick,only marks,mark=triangle,mark options={yshift=(\pgfplotmarksize+\pgflinewidth)/2}] table [y=Mbid1,x=T] from \ttfcmsh;
\addplot [color=mydarkred,thick,only marks,mark=triangle,mark options={yshift=(\pgflinewidth-\pgfplotmarksize)/2,style={rotate=180}}] table [y=Mask1,x=T] from \ttfcmsh;
\end{axis}
\end{tikzpicture}}
\hspace*{0.25cm}
\scalebox{0.9}{%
\begin{tikzpicture}
\begin{axis}[xlabel=Maturity,
                    ylabel=Futures Price,
                    ylabel style={overlay},
                    xmin=-0.1, xmax=5.1,
                    xtick={0.085,1.085,2.085,3.085,4.088,5.088},
                    xticklabels={{\tt JAN22},{\tt JAN23},{\tt JAN24},{\tt JAN25},{\tt JAN26},{\tt JAN27}},
                    ymin=7, ymax=133,
                    grid=major,
                    legend style={legend pos=south west},
                    axis background/.style={fill=gray!10}]
\addplot [color=mydarkblue,thick,smooth] table [y=Fstag3,x=T] from \ttfcmsh;
\addplot [color=mydarkblue,thick,smooth,style=dotted] table [y=Fplain3,x=T] from \ttfcmsh;
\addplot [color=mydarkmagenta,thick,smooth] table [y=Fbenth3,x=T] from \ttfcmsh;
\addplot [color=mydarkred,thick,only marks,mark=triangle,mark options={yshift=(\pgfplotmarksize+\pgflinewidth)/2}] table [y=Mbid3,x=T] from \ttfcmsh;
\addplot [color=mydarkred,thick,only marks,mark=triangle,mark options={yshift=(\pgflinewidth-\pgfplotmarksize)/2,style={rotate=180}}] table [y=Mask3,x=T] from \ttfcmsh;
\end{axis}
\end{tikzpicture}}
\\\vspace*{0.5cm}
\scalebox{0.9}{%
\begin{tikzpicture}
\begin{axis}[xlabel=Maturity,
                    ylabel=Futures Price,
                    ylabel style={overlay},
                    xmin=-0.1, xmax=5.1,
                    xtick={0.085,1.085,2.085,3.085,4.088,5.088},
                    xticklabels={{\tt JAN22},{\tt JAN23},{\tt JAN24},{\tt JAN25},{\tt JAN26},{\tt JAN27}},
                    ymin=7, ymax=133,
                    grid=major,
                    legend style={legend pos=south west},
                    axis background/.style={fill=gray!10}]
\addplot [color=mydarkblue,thick,smooth] table [y=Fstag6,x=T] from \ttfcmsh;
\addplot [color=mydarkblue,thick,smooth,style=dotted] table [y=Fplain6,x=T] from \ttfcmsh;
\addplot [color=mydarkmagenta,thick,smooth] table [y=Fbenth6,x=T] from \ttfcmsh;
\addplot [color=mydarkred,thick,only marks,mark=triangle,mark options={yshift=(\pgfplotmarksize+\pgflinewidth)/2}] table [y=Mbid6,x=T] from \ttfcmsh;
\addplot [color=mydarkred,thick,only marks,mark=triangle,mark options={yshift=(\pgflinewidth-\pgfplotmarksize)/2,style={rotate=180}}] table [y=Mask6,x=T] from \ttfcmsh;
\end{axis}
\end{tikzpicture}}
\hspace*{0.25cm}
\scalebox{0.9}{%
\begin{tikzpicture}
\begin{axis}[xlabel=Maturity,
                    ylabel=Futures Price,
                    ylabel style={overlay},
                    xmin=-0.1, xmax=5.1,
                    xtick={0.085,1.085,2.085,3.085,4.088,5.088},
                    xticklabels={{\tt JAN22},{\tt JAN23},{\tt JAN24},{\tt JAN25},{\tt JAN26},{\tt JAN27}},
                    ymin=7, ymax=133,
                    grid=major,
                    legend style={legend pos=south west},
                    axis background/.style={fill=gray!10}]
\addplot [color=mydarkblue,thick,smooth] table [y=Fstag12,x=T] from \ttfcmsh;
\addplot [color=mydarkblue,thick,smooth,style=dotted] table [y=Fplain12,x=T] from \ttfcmsh;
\addplot [color=mydarkmagenta,thick,smooth] table [y=Fbenth12,x=T] from \ttfcmsh;
\addplot [color=mydarkred,thick,only marks,mark=triangle,mark options={yshift=(\pgfplotmarksize+\pgflinewidth)/2}] table [y=Mbid12,x=T] from \ttfcmsh;
\addplot [color=mydarkred,thick,only marks,mark=triangle,mark options={yshift=(\pgflinewidth-\pgfplotmarksize)/2,style={rotate=180}}] table [y=Mask12,x=T] from \ttfcmsh;
\end{axis}
\end{tikzpicture}}
\end{center}
\caption{Term structures of futures prices for TTF natural gas quoted on 15 December 2021 on ICE market. From top-left to bottom-right the panels show the delivery periods of one, three, six months and one year. Triangles represent bid-ask futures contracts prices. Continuous blue lines is our model with penalization, dotted blue lines without it, while continuous pink lines is classical Kriging.}
\label{fig:ttfc}
\end{figure}

\begin{table}
\begin{center}
\pgfplotstabletypeset[
    col sep=tab,
    columns={id,bid,ask,Fplain,Fstag,Fbenth},
    columns/id/.style={column name=Contract, column type={c|}, string type},
    columns/bid/.style={column name=$q^b$, fixed, fixed zerofill, precision=2, dec sep align, clear infinite},
    columns/ask/.style={column name=$q^a$, fixed, fixed zerofill, precision=2, dec sep align, clear infinite},
    columns/Fplain/.style={column name=$F$, fixed, fixed zerofill, precision=2, dec sep align, clear infinite},
    columns/Fstag/.style={column name=$F^{\rm K}$, fixed, fixed zerofill, precision=2, dec sep align, clear infinite},
    columns/Fbenth/.style={column name=$F^{\rm B}$, fixed, fixed zerofill, precision=2, dec sep align, clear infinite},
    every row no 11/.style={before row=\midrule},
    every row no 27/.style={before row=\midrule},
    every row no 35/.style={before row=\midrule},
    every head row/.style={before row=\toprule,after row=\midrule},
    every last row/.style={after row=\bottomrule},
]{\ttfdmkt}
\end{center}
\caption{Market data for TTF natural gas observed on ICE exchange on 11 March 2021. From top to bottom: one-month contracts, quarters (three-month contracts), seasons (summer is the average of the second and third quarter, winter is the average of the fourth quarter and the first one of the following year), years. The columns report from left to right contract identifier, bid and ask prices, model implied prices with and without seasonality management, classical Kriging implied prices.}
\label{tab:ttfd}
\end{table}

\begin{figure}
\begin{center}
\scalebox{0.9}{%
\begin{tikzpicture}
\begin{axis}[xlabel=Maturity,
                    ylabel=Futures Price,
                    ylabel style={overlay},
                    xmin=-0.1, xmax=5.1,
                    xtick={0.838,1.838,2.841,3.841,4.841},
                    xticklabels={{\tt JAN23},{\tt JAN24},{\tt JAN25},{\tt JAN26},{\tt JAN27}},
                    ymin=7, ymax=133,
                    grid=major,
                    legend style={legend pos=south west},
                    axis background/.style={fill=gray!10}]
\addplot [color=mydarkblue,thick,smooth] table [y=Fstag1,x=T] from \ttfdmsh;
\addplot [color=mydarkblue,thick,smooth,style=dotted] table [y=Fplain1,x=T] from \ttfdmsh;
\addplot [color=mydarkmagenta,thick,smooth] table [y=Fbenth1,x=T] from \ttfdmsh;
\addplot [color=mydarkred,thick,only marks,mark=triangle,mark options={yshift=(\pgfplotmarksize+\pgflinewidth)/2}] table [y=Mbid1,x=T] from \ttfdmsh;
\addplot [color=mydarkred,thick,only marks,mark=triangle,mark options={yshift=(\pgflinewidth-\pgfplotmarksize)/2,style={rotate=180}}] table [y=Mask1,x=T] from \ttfdmsh;
\end{axis}
\end{tikzpicture}}
\hspace*{0.5cm}
\scalebox{0.9}{%
\begin{tikzpicture}
\begin{axis}[xlabel=Maturity,
                    ylabel=Futures Price,
                    ylabel style={overlay},
                    xmin=-0.1, xmax=5.1,
                    xtick={0.838,1.838,2.841,3.841,4.841},
                    xticklabels={{\tt JAN23},{\tt JAN24},{\tt JAN25},{\tt JAN26},{\tt JAN27}},
                    ymin=7, ymax=133,
                    grid=major,
                    legend style={legend pos=south west},
                    axis background/.style={fill=gray!10}]
\addplot [color=mydarkblue,thick,smooth] table [y=Fstag3,x=T] from \ttfdmsh;
\addplot [color=mydarkblue,thick,smooth,style=dotted] table [y=Fplain3,x=T] from \ttfdmsh;
\addplot [color=mydarkmagenta,thick,smooth] table [y=Fbenth3,x=T] from \ttfdmsh;
\addplot [color=mydarkred,thick,only marks,mark=triangle,mark options={yshift=(\pgfplotmarksize+\pgflinewidth)/2}] table [y=Mbid3,x=T] from \ttfdmsh;
\addplot [color=mydarkred,thick,only marks,mark=triangle,mark options={yshift=(\pgflinewidth-\pgfplotmarksize)/2,style={rotate=180}}] table [y=Mask3,x=T] from \ttfdmsh;
\end{axis}
\end{tikzpicture}}
\\\vspace*{0.5cm}
\scalebox{0.9}{%
\begin{tikzpicture}
\begin{axis}[xlabel=Maturity,
                    ylabel=Futures Price,
                    ylabel style={overlay},
                    xmin=-0.1, xmax=5.1,
                    xtick={0.838,1.838,2.841,3.841,4.841},
                    xticklabels={{\tt JAN23},{\tt JAN24},{\tt JAN25},{\tt JAN26},{\tt JAN27}},
                    ymin=7, ymax=133,
                    grid=major,
                    legend style={legend pos=south west},
                    axis background/.style={fill=gray!10}]
\addplot [color=mydarkblue,thick,smooth] table [y=Fstag6,x=T] from \ttfdmsh;
\addplot [color=mydarkblue,thick,smooth,style=dotted] table [y=Fplain6,x=T] from \ttfdmsh;
\addplot [color=mydarkmagenta,thick,smooth] table [y=Fbenth6,x=T] from \ttfdmsh;
\addplot [color=mydarkred,thick,only marks,mark=triangle,mark options={yshift=(\pgfplotmarksize+\pgflinewidth)/2}] table [y=Mbid6,x=T] from \ttfdmsh;
\addplot [color=mydarkred,thick,only marks,mark=triangle,mark options={yshift=(\pgflinewidth-\pgfplotmarksize)/2,style={rotate=180}}] table [y=Mask6,x=T] from \ttfdmsh;
\end{axis}
\end{tikzpicture}}
\hspace*{0.5cm}
\scalebox{0.9}{%
\begin{tikzpicture}
\begin{axis}[xlabel=Maturity,
                    ylabel=Futures Price,
                    ylabel style={overlay},
                    xmin=-0.1, xmax=5.1,
                    xtick={0.838,1.838,2.841,3.841,4.841},
                    xticklabels={{\tt JAN23},{\tt JAN24},{\tt JAN25},{\tt JAN26},{\tt JAN27}},
                    ymin=7, ymax=133,
                    grid=major,
                    legend style={legend pos=south west},
                    axis background/.style={fill=gray!10}]
\addplot [color=mydarkblue,thick,smooth] table [y=Fstag12,x=T] from \ttfdmsh;
\addplot [color=mydarkblue,thick,smooth,style=dotted] table [y=Fplain12,x=T] from \ttfdmsh;
\addplot [color=mydarkmagenta,thick,smooth] table [y=Fbenth12,x=T] from \ttfdmsh;
\addplot [color=mydarkred,thick,only marks,mark=triangle,mark options={yshift=(\pgfplotmarksize+\pgflinewidth)/2}] table [y=Mbid12,x=T] from \ttfdmsh;
\addplot [color=mydarkred,thick,only marks,mark=triangle,mark options={yshift=(\pgflinewidth-\pgfplotmarksize)/2,style={rotate=180}}] table [y=Mask12,x=T] from \ttfdmsh;
\end{axis}
\end{tikzpicture}}
\end{center}
\caption{Term structures of futures prices for TTF natural gas quoted on 11 March 2022 on ICE market. From top-left to bottom-right the panels show the delivery periods of one, three, six months and one year. Triangles represent bid-ask futures contracts prices. Continuous blue lines is our model with penalization, dotted blue lines without it, while continuous pink lines is classical Kriging.}
\label{fig:ttfd}
\end{figure}

\begin{table}
\begin{center}
\pgfplotstabletypeset[
    col sep=tab,
    columns={id,bid,ask,Fplain,Fstag,Fbenth},
    columns/id/.style={column name=Contract, column type={c|}, string type},
    columns/bid/.style={column name=$q^b$, fixed, fixed zerofill, precision=2, dec sep align, clear infinite},
    columns/ask/.style={column name=$q^a$, fixed, fixed zerofill, precision=2, dec sep align, clear infinite},
    columns/Fplain/.style={column name=$F$, fixed, fixed zerofill, precision=2, dec sep align, clear infinite},
    columns/Fstag/.style={column name=$F^{\rm K}$, fixed, fixed zerofill, precision=2, dec sep align, clear infinite},
    columns/Fbenth/.style={column name=$F^{\rm B}$, fixed, fixed zerofill, precision=2, dec sep align, clear infinite},
    every row no 10/.style={before row=\midrule},
    every row no 14/.style={before row=\midrule},
    every head row/.style={before row=\toprule,after row=\midrule},
    every last row/.style={after row=\bottomrule},
]{\nbpmkt}
\end{center}
\caption{Market data for NBP natural gas observed on ICE exchange on 11 March 2021. From top to bottom: one-month contracts, quarters (three-month contracts), seasons (summer is the average of the second and third quarter, winter is the average of the fourth quarter and the first one of the following year). The columns report from left to right contract identifier, bid and ask prices, model implied prices with and without seasonality management, classical Kriging implied prices.}
\label{tab:nbp}
\end{table}

\begin{figure}
\begin{center}
\scalebox{0.9}{%
\begin{tikzpicture}
\begin{axis}[xlabel=Maturity,
                    ylabel=Futures Price,
                    ylabel style={overlay},
                    xmin=-0.1, xmax=5.1,
                    xtick={0.838,1.838,2.841,3.841,4.841},
                    xticklabels={{\tt JAN23},{\tt JAN24},{\tt JAN25},{\tt JAN26},{\tt JAN27}},
                    ymin=3, ymax=347,
                    grid=major,
                    legend style={legend pos=south west},
                    axis background/.style={fill=gray!10}]
\addplot [color=mydarkblue,thick,smooth] table [y=Fstag1,x=T] from \nbpmsh;
\addplot [color=mydarkblue,thick,smooth,style=dotted] table [y=Fplain1,x=T] from \nbpmsh;
\addplot [color=mydarkmagenta,thick,smooth] table [y=Fbenth1,x=T] from \nbpmsh;
\addplot [color=mydarkred,thick,only marks,mark=triangle,mark options={yshift=(\pgfplotmarksize+\pgflinewidth)/2}] table [y=Mbid1,x=T] from \nbpmsh;
\addplot [color=mydarkred,thick,only marks,mark=triangle,mark options={yshift=(\pgflinewidth-\pgfplotmarksize)/2,style={rotate=180}}] table [y=Mask1,x=T] from \nbpmsh;
\end{axis}
\end{tikzpicture}}
\hspace*{0.5cm}
\scalebox{0.9}{%
\begin{tikzpicture}
\begin{axis}[xlabel=Maturity,
                    ylabel=Futures Price,
                    ylabel style={overlay},
                    xmin=-0.1, xmax=5.1,
                    xtick={0.838,1.838,2.841,3.841,4.841},
                    xticklabels={{\tt JAN23},{\tt JAN24},{\tt JAN25},{\tt JAN26},{\tt JAN27}},
                    ymin=3, ymax=347,
                    grid=major,
                    legend style={legend pos=south west},
                    axis background/.style={fill=gray!10}]
\addplot [color=mydarkblue,thick,smooth] table [y=Fstag3,x=T] from \nbpmsh;
\addplot [color=mydarkblue,thick,smooth,style=dotted] table [y=Fplain3,x=T] from \nbpmsh;
\addplot [color=mydarkmagenta,thick,smooth] table [y=Fbenth3,x=T] from \nbpmsh;
\addplot [color=mydarkred,thick,only marks,mark=triangle,mark options={yshift=(\pgfplotmarksize+\pgflinewidth)/2}] table [y=Mbid3,x=T] from \nbpmsh;
\addplot [color=mydarkred,thick,only marks,mark=triangle,mark options={yshift=(\pgflinewidth-\pgfplotmarksize)/2,style={rotate=180}}] table [y=Mask3,x=T] from \nbpmsh;
\end{axis}
\end{tikzpicture}}
\\\vspace*{0.5cm}
\scalebox{0.9}{%
\begin{tikzpicture}
\begin{axis}[xlabel=Maturity,
                    ylabel=Futures Price,
                    ylabel style={overlay},
                    xmin=-0.1, xmax=5.1,
                    xtick={0.838,1.838,2.841,3.841,4.841},
                    xticklabels={{\tt JAN23},{\tt JAN24},{\tt JAN25},{\tt JAN26},{\tt JAN27}},
                    ymin=3, ymax=347,
                    grid=major,
                    legend style={legend pos=south west},
                    axis background/.style={fill=gray!10}]
\addplot [color=mydarkblue,thick,smooth] table [y=Fstag6,x=T] from \nbpmsh;
\addplot [color=mydarkblue,thick,smooth,style=dotted] table [y=Fplain6,x=T] from \nbpmsh;
\addplot [color=mydarkmagenta,thick,smooth] table [y=Fbenth6,x=T] from \nbpmsh;
\addplot [color=mydarkred,thick,only marks,mark=triangle,mark options={yshift=(\pgfplotmarksize+\pgflinewidth)/2}] table [y=Mbid6,x=T] from \nbpmsh;
\addplot [color=mydarkred,thick,only marks,mark=triangle,mark options={yshift=(\pgflinewidth-\pgfplotmarksize)/2,style={rotate=180}}] table [y=Mask6,x=T] from \nbpmsh;
\end{axis}
\end{tikzpicture}}
\hspace*{0.5cm}
\scalebox{0.9}{%
\begin{tikzpicture}
\begin{axis}[xlabel=Maturity,
                    ylabel=Futures Price,
                    ylabel style={overlay},
                    xmin=-0.1, xmax=5.1,
                    xtick={0.838,1.838,2.841,3.841,4.841},
                    xticklabels={{\tt JAN23},{\tt JAN24},{\tt JAN25},{\tt JAN26},{\tt JAN27}},
                    ymin=3, ymax=347,
                    grid=major,
                    legend style={legend pos=south west},
                    axis background/.style={fill=gray!10}]
\addplot [color=mydarkblue,thick,smooth] table [y=Fstag12,x=T] from \nbpmsh;
\addplot [color=mydarkblue,thick,smooth,style=dotted] table [y=Fplain12,x=T] from \nbpmsh;
\addplot [color=mydarkmagenta,thick,smooth] table [y=Fbenth12,x=T] from \nbpmsh;
\addplot [color=mydarkred,thick,only marks,mark=triangle,mark options={yshift=(\pgfplotmarksize+\pgflinewidth)/2}] table [y=Mbid12,x=T] from \nbpmsh;
\addplot [color=mydarkred,thick,only marks,mark=triangle,mark options={yshift=(\pgflinewidth-\pgfplotmarksize)/2,style={rotate=180}}] table [y=Mask12,x=T] from \nbpmsh;
\end{axis}
\end{tikzpicture}}
\end{center}
\caption{Term structures of futures prices for NBP natural gas quoted on 11 March 2022 on ICE market. From top-left to bottom-right the panels show the delivery periods of one, three, six months and one year. Triangles represent bid-ask futures contracts prices. Continuous blue lines is our model with penalization, dotted blue lines without it, while continuous pink lines is classical Kriging.}
\label{fig:nbp}
\end{figure}

\begin{table}
\begin{center}
\pgfplotstabletypeset[
    col sep=tab,
    columns={id,bid,ask,Fplain,Fstag,Fbenth},
    columns/id/.style={column name=Contract, column type={c|}, string type},
    columns/bid/.style={column name=$q^b$, fixed, fixed zerofill, precision=2, dec sep align, clear infinite},
    columns/ask/.style={column name=$q^a$, fixed, fixed zerofill, precision=2, dec sep align, clear infinite},
    columns/Fplain/.style={column name=$F$, fixed, fixed zerofill, precision=2, dec sep align, clear infinite},
    columns/Fstag/.style={column name=$F^{\rm K}$, fixed, fixed zerofill, precision=2, dec sep align, clear infinite},
    columns/Fbenth/.style={column name=$F^{\rm B}$, fixed, fixed zerofill, precision=2, dec sep align, clear infinite},
    every head row/.style={before row=\toprule,after row=\midrule},
    every last row/.style={after row=\bottomrule},
]{\ngmkt}
\end{center}
\caption{Market data for NG natural gas observed on NYMEX exchange on 11 March 2021. One-month contracts. The columns report from left to right contract identifier, bid and ask prices, model implied prices with and without seasonality management, classical Kriging implied prices.}
\label{tab:ng}
\end{table}

\begin{figure}
\begin{center}
\scalebox{0.9}{%
\begin{tikzpicture}
\begin{axis}[xlabel=Maturity,
                    ylabel=Futures Price,
                    ylabel style={overlay},
                    xmin=-0.1, xmax=5.1,
                    xtick={0.838,1.838,2.841,3.841,4.841},
                    xticklabels={{\tt JAN23},{\tt JAN24},{\tt JAN25},{\tt JAN26},{\tt JAN27}},
                    ymin=2, ymax=6,
                    grid=major,
                    legend style={legend pos=south west},
                    axis background/.style={fill=gray!10}]
\addplot [color=mydarkblue,thick,smooth] table [y=Fstag1,x=T] from \ngmsh;
\addplot [color=mydarkblue,thick,smooth,style=dotted] table [y=Fplain1,x=T] from \ngmsh;
\addplot [color=mydarkmagenta,thick,smooth] table [y=Fbenth1,x=T] from \ngmsh;
\addplot [color=mydarkred,thick,only marks,mark=triangle,mark options={yshift=(\pgfplotmarksize+\pgflinewidth)/2}] table [y=Mbid1,x=T] from \ngmsh;
\addplot [color=mydarkred,thick,only marks,mark=triangle,mark options={yshift=(\pgflinewidth-\pgfplotmarksize)/2,style={rotate=180}}] table [y=Mask1,x=T] from \ngmsh;
\end{axis}
\end{tikzpicture}}
\hspace*{0.5cm}
\scalebox{0.9}{%
\begin{tikzpicture}
\begin{axis}[xlabel=Maturity,
                    ylabel=Futures Price,
                    ylabel style={overlay},
                    xmin=-0.1, xmax=5.1,
                    xtick={0.838,1.838,2.841,3.841,4.841},
                    xticklabels={{\tt JAN23},{\tt JAN24},{\tt JAN25},{\tt JAN26},{\tt JAN27}},
                    ymin=2, ymax=6,
                    grid=major,
                    legend style={legend pos=south west},
                    axis background/.style={fill=gray!10}]
\addplot [color=mydarkblue,thick,smooth] table [y=Fstag3,x=T] from \ngmsh;
\addplot [color=mydarkblue,thick,smooth,style=dotted] table [y=Fplain3,x=T] from \ngmsh;
\addplot [color=mydarkmagenta,thick,smooth] table [y=Fbenth3,x=T] from \ngmsh;
\addplot [color=mydarkred,thick,only marks,mark=triangle,mark options={yshift=(\pgfplotmarksize+\pgflinewidth)/2}] table [y=Mbid3,x=T] from \ngmsh;
\addplot [color=mydarkred,thick,only marks,mark=triangle,mark options={yshift=(\pgflinewidth-\pgfplotmarksize)/2,style={rotate=180}}] table [y=Mask3,x=T] from \ngmsh;
\end{axis}
\end{tikzpicture}}
\\\vspace*{0.5cm}
\scalebox{0.9}{%
\begin{tikzpicture}
\begin{axis}[xlabel=Maturity,
                    ylabel=Futures Price,
                    ylabel style={overlay},
                    xmin=-0.1, xmax=5.1,
                    xtick={0.838,1.838,2.841,3.841,4.841},
                    xticklabels={{\tt JAN23},{\tt JAN24},{\tt JAN25},{\tt JAN26},{\tt JAN27}},
                    ymin=2, ymax=6,
                    grid=major,
                    legend style={legend pos=south west},
                    axis background/.style={fill=gray!10}]
\addplot [color=mydarkblue,thick,smooth] table [y=Fstag6,x=T] from \ngmsh;
\addplot [color=mydarkblue,thick,smooth,style=dotted] table [y=Fplain6,x=T] from \ngmsh;
\addplot [color=mydarkmagenta,thick,smooth] table [y=Fbenth6,x=T] from \ngmsh;
\addplot [color=mydarkred,thick,only marks,mark=triangle,mark options={yshift=(\pgfplotmarksize+\pgflinewidth)/2}] table [y=Mbid6,x=T] from \ngmsh;
\addplot [color=mydarkred,thick,only marks,mark=triangle,mark options={yshift=(\pgflinewidth-\pgfplotmarksize)/2,style={rotate=180}}] table [y=Mask6,x=T] from \ngmsh;
\end{axis}
\end{tikzpicture}}
\hspace*{0.5cm}
\scalebox{0.9}{%
\begin{tikzpicture}
\begin{axis}[xlabel=Maturity,
                    ylabel=Futures Price,
                    ylabel style={overlay},
                    xmin=-0.1, xmax=5.1,
                    xtick={0.838,1.838,2.841,3.841,4.841},
                    xticklabels={{\tt JAN23},{\tt JAN24},{\tt JAN25},{\tt JAN26},{\tt JAN27}},
                    ymin=2, ymax=6,
                    grid=major,
                    legend style={legend pos=south west},
                    axis background/.style={fill=gray!10}]
\addplot [color=mydarkblue,thick,smooth] table [y=Fstag12,x=T] from \ngmsh;
\addplot [color=mydarkblue,thick,smooth,style=dotted] table [y=Fplain12,x=T] from \ngmsh;
\addplot [color=mydarkmagenta,thick,smooth] table [y=Fbenth12,x=T] from \ngmsh;
\addplot [color=mydarkred,thick,only marks,mark=triangle,mark options={yshift=(\pgfplotmarksize+\pgflinewidth)/2}] table [y=Mbid12,x=T] from \ngmsh;
\addplot [color=mydarkred,thick,only marks,mark=triangle,mark options={yshift=(\pgflinewidth-\pgfplotmarksize)/2,style={rotate=180}}] table [y=Mask12,x=T] from \ngmsh;
\end{axis}
\end{tikzpicture}}
\end{center}
\caption{Term structures of futures prices for NG natural gas quoted on 11 March 2022 on NYMEX market. From top-left to bottom-right the panels show the delivery periods of one, three, six months and one year. Triangles represent bid-ask futures contracts prices. Continuous blue lines is our model with penalization, dotted blue lines without it, while continuous pink lines is classical Kriging.}
\label{fig:ng}
\end{figure}

\subsection{Electricity}

We investigate electricity prices on the German grid. In particular, we consider the base-load from EEX on 15 December 2021 (table \ref{tab:powerde}).

We notice that the market can quote futures with different delivery periods (month, quarter, and year). We can see a bi-modal yearly seasonal pattern, a smaller peak in summer followed by a bigger peak in winter, superimposed on a small decreasing trend (figure \ref{fig:powerde}). Our model integrating the information from all the market delivery periods is able to reproduce the seasonal pattern of the first year in the following ones. On the contrary, classical Kriging, which can be calibrated only to one-month quotes, fails to show a reasonable behaviour for longer maturities. We notice also that the penalty term introduced in equation \eqref{eq:estimate-seasonality} is here fundamental to catch the complex behaviour of the seasonal pattern.

\begin{table}
\begin{center}
\pgfplotstabletypeset[
    col sep=tab,
    columns={id,bid,ask,Fplain,Fstag,Fbenth},
    columns/id/.style={column name=Contract, column type={c|}, string type},
    columns/bid/.style={column name=$q^b$, fixed, fixed zerofill, precision=2, dec sep align, clear infinite},
    columns/ask/.style={column name=$q^a$, fixed, fixed zerofill, precision=2, dec sep align, clear infinite},
    columns/Fplain/.style={column name=$F$, fixed, fixed zerofill, precision=2, dec sep align, clear infinite},
    columns/Fstag/.style={column name=$F^{\rm K}$, fixed, fixed zerofill, precision=2, dec sep align, clear infinite},
    columns/Fbenth/.style={column name=$F^{\rm B}$, fixed, fixed zerofill, precision=2, dec sep align, clear infinite},
    every row no 3/.style={before row=\midrule},
    every row no 8/.style={before row=\midrule},
    every head row/.style={before row=\toprule,after row=\midrule},
    every last row/.style={after row=\bottomrule},
]{\germanypowermkt}
\end{center}
\caption{Market data for German power observed on EEX exchange on 15 December 2021. From top to bottom: one-month contracts, quarters (three-month contracts), years. The columns report from left to right contract identifier, bid and ask prices, model implied prices with and without seasonality management, classical Kriging implied prices.}
\label{tab:powerde}
\end{table}

\begin{figure}
\begin{center}
\scalebox{0.9}{%
\begin{tikzpicture}
\begin{axis}[xlabel=Maturity,
                    ylabel=Futures Price,
                    ylabel style={overlay},
                    xmin=-0.1, xmax=3.1,
                    xtick={0.085,1.085,2.085,3.085},
                    xticklabels={{\tt JAN22},{\tt JAN23},{\tt JAN24},{\tt JAN25}},
                    ymin=7, ymax=333,
                    grid=major,
                    legend style={legend pos=south west},
                    axis background/.style={fill=gray!10}]
\addplot [color=mydarkblue,thick,smooth] table [y=Fstag1,x=T] from \germanypowermsh;
\addplot [color=mydarkblue,thick,smooth,style=dotted] table [y=Fplain1,x=T] from \germanypowermsh;
\addplot [color=mydarkmagenta,thick,smooth] table [y=Fbenth1,x=T] from \germanypowermsh;
\addplot [color=mydarkred,thick,only marks,mark=triangle,mark options={yshift=(\pgfplotmarksize+\pgflinewidth)/2}] table [y=Mbid1,x=T] from \germanypowermsh;
\addplot [color=mydarkred,thick,only marks,mark=triangle,mark options={yshift=(\pgflinewidth-\pgfplotmarksize)/2,style={rotate=180}}] table [y=Mask1,x=T] from \germanypowermsh;
\end{axis}
\end{tikzpicture}}
\hspace*{0.25cm}
\scalebox{0.9}{%
\begin{tikzpicture}
\begin{axis}[xlabel=Maturity,
                    ylabel=Futures Price,
                    ylabel style={overlay},
                    xmin=-0.1, xmax=3.1,
                    xtick={0.085,1.085,2.085,3.085},
                    xticklabels={{\tt JAN22},{\tt JAN23},{\tt JAN24},{\tt JAN25}},
                    ymin=7, ymax=333,
                    grid=major,
                    legend style={legend pos=south west},
                    axis background/.style={fill=gray!10}]
\addplot [color=mydarkblue,thick,smooth] table [y=Fstag3,x=T] from \germanypowermsh;
\addplot [color=mydarkblue,thick,smooth,style=dotted] table [y=Fplain3,x=T] from \germanypowermsh;
\addplot [color=mydarkmagenta,thick,smooth] table [y=Fbenth3,x=T] from \germanypowermsh;
\addplot [color=mydarkred,thick,only marks,mark=triangle,mark options={yshift=(\pgfplotmarksize+\pgflinewidth)/2}] table [y=Mbid3,x=T] from \germanypowermsh;
\addplot [color=mydarkred,thick,only marks,mark=triangle,mark options={yshift=(\pgflinewidth-\pgfplotmarksize)/2,style={rotate=180}}] table [y=Mask3,x=T] from \germanypowermsh;
\end{axis}
\end{tikzpicture}}
\\\vspace*{0.5cm}
\scalebox{0.9}{%
\begin{tikzpicture}
\begin{axis}[xlabel=Maturity,
                    ylabel=Futures Price,
                    ylabel style={overlay},
                    xmin=-0.1, xmax=3.1,
                    xtick={0.085,1.085,2.085,3.085},
                    xticklabels={{\tt JAN22},{\tt JAN23},{\tt JAN24},{\tt JAN25}},
                    ymin=7, ymax=333,
                    grid=major,
                    legend style={legend pos=south west},
                    axis background/.style={fill=gray!10}]
\addplot [color=mydarkblue,thick,smooth] table [y=Fstag6,x=T] from \germanypowermsh;
\addplot [color=mydarkblue,thick,smooth,style=dotted] table [y=Fplain6,x=T] from \germanypowermsh;
\addplot [color=mydarkmagenta,thick,smooth] table [y=Fbenth6,x=T] from \germanypowermsh;
\addplot [color=mydarkred,thick,only marks,mark=triangle,mark options={yshift=(\pgfplotmarksize+\pgflinewidth)/2}] table [y=Mbid6,x=T] from \germanypowermsh;
\addplot [color=mydarkred,thick,only marks,mark=triangle,mark options={yshift=(\pgflinewidth-\pgfplotmarksize)/2,style={rotate=180}}] table [y=Mask6,x=T] from \germanypowermsh;
\end{axis}
\end{tikzpicture}}
\hspace*{0.25cm}
\scalebox{0.9}{%
\begin{tikzpicture}
\begin{axis}[xlabel=Maturity,
                    ylabel=Futures Price,
                    ylabel style={overlay},
                    xmin=-0.1, xmax=3.1,
                    xtick={0.085,1.085,2.085,3.085},
                    xticklabels={{\tt JAN22},{\tt JAN23},{\tt JAN24},{\tt JAN25}},
                    ymin=7, ymax=333,
                    grid=major,
                    legend style={legend pos=south west},
                    axis background/.style={fill=gray!10}]
\addplot [color=mydarkblue,thick,smooth] table [y=Fstag12,x=T] from \germanypowermsh;
\addplot [color=mydarkblue,thick,smooth,style=dotted] table [y=Fplain12,x=T] from \germanypowermsh;
\addplot [color=mydarkmagenta,thick,smooth] table [y=Fbenth12,x=T] from \germanypowermsh;
\addplot [color=mydarkred,thick,only marks,mark=triangle,mark options={yshift=(\pgfplotmarksize+\pgflinewidth)/2}] table [y=Mbid12,x=T] from \germanypowermsh;
\addplot [color=mydarkred,thick,only marks,mark=triangle,mark options={yshift=(\pgflinewidth-\pgfplotmarksize)/2,style={rotate=180}}] table [y=Mask12,x=T] from \germanypowermsh;
\end{axis}
\end{tikzpicture}}
\end{center}
\caption{Term structures of futures prices for German power quoted on 15 December 2021 on EEX market. From top-left to bottom-right the panels show the delivery periods of one, three, six months and one year. Triangles represent bid-ask futures contracts prices. Continuous blue lines is our model with penalization, dotted blue lines without it, while continuous pink lines is classical Kriging.}
\label{fig:powerde}
\end{figure}

\subsection{Oils}

We investigate various oil prices on different observation dates and different trading points. In particular, we consider WTI from NYMEX on 15 December 2021 (table \ref{tab:wti}), Brent from ICE on 15 December 2021 (table \ref{tab:brent}), heating oil from NYMEX on 15 December 2021 (table \ref{tab:heatingoila}) and 11 March 2022 (table \ref{tab:heatingoilb}).

We notice that the market quotes only monthly futures. If we analyze data for WTI and Brent we can see a stable decreasing trend well recovered both by our model (with and without penalty terms for seasonal patterns) and by classical Kriging (figures \ref{fig:wti} and \ref{fig:brent}). A similar bahaviour is displayed also by heating oil prices on both the observation dates we considered, although we can see that classical Kriging presents a strange change in convexity in extrapolation.

\begin{table}
\begin{center}
\pgfplotstabletypeset[
    col sep=tab,
    columns={id,bid,ask,Fplain,Fstag,Fbenth},
    columns/id/.style={column name=Contract, column type={c|}, string type},
    columns/bid/.style={column name=$q^b$, fixed, fixed zerofill, precision=2, dec sep align, clear infinite},
    columns/ask/.style={column name=$q^a$, fixed, fixed zerofill, precision=2, dec sep align, clear infinite},
    columns/Fplain/.style={column name=$F$, fixed, fixed zerofill, precision=2, dec sep align, clear infinite},
    columns/Fstag/.style={column name=$F^{\rm K}$, fixed, fixed zerofill, precision=2, dec sep align, clear infinite},
    columns/Fbenth/.style={column name=$F^{\rm B}$, fixed, fixed zerofill, precision=2, dec sep align, clear infinite},
    every head row/.style={before row=\toprule,after row=\midrule},
    every last row/.style={after row=\bottomrule},
]{\wtimkt}
\end{center}
\caption{Market data for WTI crude oil observed on NYMEX exchange on 15 December 2021. One-month contracts. The columns report from left to right contract identifier, bid and ask prices, model implied prices with and without seasonality management, classical Kriging implied prices.}
\label{tab:wti}
\end{table}

\begin{figure}
\begin{center}
\scalebox{0.9}{%
\begin{tikzpicture}
\begin{axis}[xlabel=Maturity,
                    ylabel=Futures Price,
                    ylabel style={overlay},
                    xmin=-0.1, xmax=6.1,
                    xtick={0.000,1.00,2.000,3.003,4.003,5.003,6.003},
                    xticklabels={{\tt JAN22},{\tt JAN23},{\tt JAN24},{\tt JAN25},{\tt JAN26},{\tt JAN27},{\tt JAN28}},
                    ymin=57, ymax=77,
                    grid=major,
                    legend style={legend pos=south west},
                    axis background/.style={fill=gray!10}]
\addplot [color=mydarkblue,thick,smooth] table [y=Fstag1,x=T] from \wtimsh;
\addplot [color=mydarkblue,thick,smooth,style=dotted] table [y=Fplain1,x=T] from \wtimsh;
\addplot [color=mydarkmagenta,thick,smooth] table [y=Fbenth1,x=T] from \wtimsh;
\addplot [color=mydarkred,thick,only marks,mark=triangle,mark options={yshift=(\pgfplotmarksize+\pgflinewidth)/2}] table [y=Mbid1,x=T] from \wtimsh;
\addplot [color=mydarkred,thick,only marks,mark=triangle,mark options={yshift=(\pgflinewidth-\pgfplotmarksize)/2,style={rotate=180}}] table [y=Mask1,x=T] from \wtimsh;
\end{axis}
\end{tikzpicture}}
\hspace*{0.5cm}
\scalebox{0.9}{%
\begin{tikzpicture}
\begin{axis}[xlabel=Maturity,
                    ylabel=Futures Price,
                    ylabel style={overlay},
                    xmin=-0.1, xmax=6.1,
                    xtick={0.000,1.00,2.000,3.003,4.003,5.003,6.003},
                    xticklabels={{\tt JAN22},{\tt JAN23},{\tt JAN24},{\tt JAN25},{\tt JAN26},{\tt JAN27},{\tt JAN28}},
                    ymin=57, ymax=77,
                    grid=major,
                    legend style={legend pos=south west},
                    axis background/.style={fill=gray!10}]
\addplot [color=mydarkblue,thick,smooth] table [y=Fstag3,x=T] from \wtimsh;
\addplot [color=mydarkblue,thick,smooth,style=dotted] table [y=Fplain3,x=T] from \wtimsh;
\addplot [color=mydarkmagenta,thick,smooth] table [y=Fbenth3,x=T] from \wtimsh;
\addplot [color=mydarkred,thick,only marks,mark=triangle,mark options={yshift=(\pgfplotmarksize+\pgflinewidth)/2}] table [y=Mbid3,x=T] from \wtimsh;
\addplot [color=mydarkred,thick,only marks,mark=triangle,mark options={yshift=(\pgflinewidth-\pgfplotmarksize)/2,style={rotate=180}}] table [y=Mask3,x=T] from \wtimsh;
\end{axis}
\end{tikzpicture}}
\\\vspace*{0.5cm}
\scalebox{0.9}{%
\begin{tikzpicture}
\begin{axis}[xlabel=Maturity,
                    ylabel=Futures Price,
                    ylabel style={overlay},
                    xmin=-0.1, xmax=6.1,
                    xtick={0.000,1.00,2.000,3.003,4.003,5.003,6.003},
                    xticklabels={{\tt JAN22},{\tt JAN23},{\tt JAN24},{\tt JAN25},{\tt JAN26},{\tt JAN27},{\tt JAN28}},
                    ymin=57, ymax=77,
                    grid=major,
                    legend style={legend pos=south west},
                    axis background/.style={fill=gray!10}]
\addplot [color=mydarkblue,thick,smooth] table [y=Fstag6,x=T] from \wtimsh;
\addplot [color=mydarkblue,thick,smooth,style=dotted] table [y=Fplain6,x=T] from \wtimsh;
\addplot [color=mydarkmagenta,thick,smooth] table [y=Fbenth6,x=T] from \wtimsh;
\addplot [color=mydarkred,thick,only marks,mark=triangle,mark options={yshift=(\pgfplotmarksize+\pgflinewidth)/2}] table [y=Mbid6,x=T] from \wtimsh;
\addplot [color=mydarkred,thick,only marks,mark=triangle,mark options={yshift=(\pgflinewidth-\pgfplotmarksize)/2,style={rotate=180}}] table [y=Mask6,x=T] from \wtimsh;
\end{axis}
\end{tikzpicture}}
\hspace*{0.5cm}
\scalebox{0.9}{%
\begin{tikzpicture}
\begin{axis}[xlabel=Maturity,
                    ylabel=Futures Price,
                    ylabel style={overlay},
                    xmin=-0.1, xmax=6.1,
                    xtick={0.000,1.00,2.000,3.003,4.003,5.003,6.003},
                    xticklabels={{\tt JAN22},{\tt JAN23},{\tt JAN24},{\tt JAN25},{\tt JAN26},{\tt JAN27},{\tt JAN28}},
                    ymin=57, ymax=77,
                    grid=major,
                    legend style={legend pos=south west},
                    axis background/.style={fill=gray!10}]
\addplot [color=mydarkblue,thick,smooth] table [y=Fstag12,x=T] from \wtimsh;
\addplot [color=mydarkblue,thick,smooth,style=dotted] table [y=Fplain12,x=T] from \wtimsh;
\addplot [color=mydarkmagenta,thick,smooth] table [y=Fbenth12,x=T] from \wtimsh;
\addplot [color=mydarkred,thick,only marks,mark=triangle,mark options={yshift=(\pgfplotmarksize+\pgflinewidth)/2}] table [y=Mbid12,x=T] from \wtimsh;
\addplot [color=mydarkred,thick,only marks,mark=triangle,mark options={yshift=(\pgflinewidth-\pgfplotmarksize)/2,style={rotate=180}}] table [y=Mask12,x=T] from \wtimsh;
\end{axis}
\end{tikzpicture}}
\end{center}
\caption{Term structures of futures prices for WTI crude oil quoted on 15 December 2021 on NYMEX market. From top-left to bottom-right the panels show the delivery periods of one, three, six months and one year. Triangles represent bid-ask futures contracts prices. Continuous blue lines is our model with penalization, dotted blue lines without it, while continuous pink lines is classical Kriging.}
\label{fig:wti}
\end{figure}

\begin{table}
\begin{center}
\pgfplotstabletypeset[
    col sep=tab,
    columns={id,bid,ask,Fplain,Fstag,Fbenth},
    columns/id/.style={column name=Contract, column type={c|}, string type},
    columns/bid/.style={column name=$q^b$, fixed, fixed zerofill, precision=2, dec sep align, clear infinite},
    columns/ask/.style={column name=$q^a$, fixed, fixed zerofill, precision=2, dec sep align, clear infinite},
    columns/Fplain/.style={column name=$F$, fixed, fixed zerofill, precision=2, dec sep align, clear infinite},
    columns/Fstag/.style={column name=$F^{\rm K}$, fixed, fixed zerofill, precision=2, dec sep align, clear infinite},
    columns/Fbenth/.style={column name=$F^{\rm B}$, fixed, fixed zerofill, precision=2, dec sep align, clear infinite},
    every head row/.style={before row=\toprule,after row=\midrule},
    every last row/.style={after row=\bottomrule},
]{\brentmkt}
\end{center}
\caption{Market data for Brent crude oil observed on ICE exchange on 15 December 2021. One-month contracts. The columns report from left to right contract identifier, bid and ask prices, model implied prices with and without seasonality management, classical Kriging implied prices.}
\label{tab:brent}
\end{table}

\begin{figure}
\begin{center}
\scalebox{0.9}{%
\begin{tikzpicture}
\begin{axis}[xlabel=Maturity,
                    ylabel=Futures Price,
                    ylabel style={overlay},
                    xmin=-0.1, xmax=6.1,
                    xtick={0.000,1.00,2.000,3.003,4.003,5.003,6.003},
                    xticklabels={{\tt JAN22},{\tt JAN23},{\tt JAN24},{\tt JAN25},{\tt JAN26},{\tt JAN27},{\tt JAN28}},
                    ymin=57, ymax=77,
                    grid=major,
                    legend style={legend pos=south west},
                    axis background/.style={fill=gray!10}]
\addplot [color=mydarkblue,thick,smooth] table [y=Fstag1,x=T] from \brentmsh;
\addplot [color=mydarkblue,thick,smooth,style=dotted] table [y=Fplain1,x=T] from \brentmsh;
\addplot [color=mydarkmagenta,thick,smooth] table [y=Fbenth1,x=T] from \brentmsh;
\addplot [color=mydarkred,thick,only marks,mark=triangle,mark options={yshift=(\pgfplotmarksize+\pgflinewidth)/2}] table [y=Mbid1,x=T] from \brentmsh;
\addplot [color=mydarkred,thick,only marks,mark=triangle,mark options={yshift=(\pgflinewidth-\pgfplotmarksize)/2,style={rotate=180}}] table [y=Mask1,x=T] from \brentmsh;
\end{axis}
\end{tikzpicture}}
\hspace*{0.5cm}
\scalebox{0.9}{%
\begin{tikzpicture}
\begin{axis}[xlabel=Maturity,
                    ylabel=Futures Price,
                    ylabel style={overlay},
                    xmin=-0.1, xmax=6.1,
                    xtick={0.000,1.00,2.000,3.003,4.003,5.003,6.003},
                    xticklabels={{\tt JAN22},{\tt JAN23},{\tt JAN24},{\tt JAN25},{\tt JAN26},{\tt JAN27},{\tt JAN28}},
                    ymin=57, ymax=77,
                    grid=major,
                    legend style={legend pos=south west},
                    axis background/.style={fill=gray!10}]
\addplot [color=mydarkblue,thick,smooth] table [y=Fstag3,x=T] from \brentmsh;
\addplot [color=mydarkblue,thick,smooth,style=dotted] table [y=Fplain3,x=T] from \brentmsh;
\addplot [color=mydarkmagenta,thick,smooth] table [y=Fbenth3,x=T] from \brentmsh;
\addplot [color=mydarkred,thick,only marks,mark=triangle,mark options={yshift=(\pgfplotmarksize+\pgflinewidth)/2}] table [y=Mbid3,x=T] from \brentmsh;
\addplot [color=mydarkred,thick,only marks,mark=triangle,mark options={yshift=(\pgflinewidth-\pgfplotmarksize)/2,style={rotate=180}}] table [y=Mask3,x=T] from \brentmsh;
\end{axis}
\end{tikzpicture}}
\\\vspace*{0.5cm}
\scalebox{0.9}{%
\begin{tikzpicture}
\begin{axis}[xlabel=Maturity,
                    ylabel=Futures Price,
                    ylabel style={overlay},
                    xmin=-0.1, xmax=6.1,
                    xtick={0.000,1.00,2.000,3.003,4.003,5.003,6.003},
                    xticklabels={{\tt JAN22},{\tt JAN23},{\tt JAN24},{\tt JAN25},{\tt JAN26},{\tt JAN27},{\tt JAN28}},
                    ymin=57, ymax=77,
                    grid=major,
                    legend style={legend pos=south west},
                    axis background/.style={fill=gray!10}]
\addplot [color=mydarkblue,thick,smooth] table [y=Fstag6,x=T] from \brentmsh;
\addplot [color=mydarkblue,thick,smooth,style=dotted] table [y=Fplain6,x=T] from \brentmsh;
\addplot [color=mydarkmagenta,thick,smooth] table [y=Fbenth6,x=T] from \brentmsh;
\addplot [color=mydarkred,thick,only marks,mark=triangle,mark options={yshift=(\pgfplotmarksize+\pgflinewidth)/2}] table [y=Mbid6,x=T] from \brentmsh;
\addplot [color=mydarkred,thick,only marks,mark=triangle,mark options={yshift=(\pgflinewidth-\pgfplotmarksize)/2,style={rotate=180}}] table [y=Mask6,x=T] from \brentmsh;
\end{axis}
\end{tikzpicture}}
\hspace*{0.5cm}
\scalebox{0.9}{%
\begin{tikzpicture}
\begin{axis}[xlabel=Maturity,
                    ylabel=Futures Price,
                    ylabel style={overlay},
                    xmin=-0.1, xmax=6.1,
                    xtick={0.000,1.00,2.000,3.003,4.003,5.003,6.003},
                    xticklabels={{\tt JAN22},{\tt JAN23},{\tt JAN24},{\tt JAN25},{\tt JAN26},{\tt JAN27},{\tt JAN28}},
                    ymin=57, ymax=77,
                    grid=major,
                    legend style={legend pos=south west},
                    axis background/.style={fill=gray!10}]
\addplot [color=mydarkblue,thick,smooth] table [y=Fstag12,x=T] from \brentmsh;
\addplot [color=mydarkblue,thick,smooth,style=dotted] table [y=Fplain12,x=T] from \brentmsh;
\addplot [color=mydarkmagenta,thick,smooth] table [y=Fbenth12,x=T] from \brentmsh;
\addplot [color=mydarkred,thick,only marks,mark=triangle,mark options={yshift=(\pgfplotmarksize+\pgflinewidth)/2}] table [y=Mbid12,x=T] from \brentmsh;
\addplot [color=mydarkred,thick,only marks,mark=triangle,mark options={yshift=(\pgflinewidth-\pgfplotmarksize)/2,style={rotate=180}}] table [y=Mask12,x=T] from \brentmsh;
\end{axis}
\end{tikzpicture}}
\end{center}
\caption{Term structures of futures prices for Brent crude oil quoted on 15 December 2021 on ICE market. From top-left to bottom-right the panels show the delivery periods of one, three, six months and one year. Triangles represent bid-ask futures contracts prices. Continuous blue lines is our model with penalization, dotted blue lines without it, while continuous pink lines is classical Kriging.}
\label{fig:brent}
\end{figure}

\begin{table}
\begin{center}
\pgfplotstabletypeset[
    col sep=tab,
    columns={id,bid,ask,Fplain,Fstag,Fbenth},
    columns/id/.style={column name=Contract, column type={c|}, string type},
    columns/bid/.style={column name=$q^b$, fixed, fixed zerofill, precision=2, dec sep align, clear infinite},
    columns/ask/.style={column name=$q^a$, fixed, fixed zerofill, precision=2, dec sep align, clear infinite},
    columns/Fplain/.style={column name=$F$, fixed, fixed zerofill, precision=2, dec sep align, clear infinite},
    columns/Fstag/.style={column name=$F^{\rm K}$, fixed, fixed zerofill, precision=2, dec sep align, clear infinite},
    columns/Fbenth/.style={column name=$F^{\rm B}$, fixed, fixed zerofill, precision=2, dec sep align, clear infinite},
    every head row/.style={before row=\toprule,after row=\midrule},
    every last row/.style={after row=\bottomrule},
]{\heatingoilamkt}
\end{center}
\caption{Market data for Heating Oil observed on NYMEX exchange on 15 December 2021. One-month contracts. The columns report from left to right contract identifier, bid and ask prices, model implied prices with and without seasonality management, classical Kriging implied prices.}
\label{tab:heatingoila}
\end{table}

\begin{figure}
\begin{center}
\scalebox{0.9}{%
\begin{tikzpicture}
\begin{axis}[xlabel=Maturity,
                    ylabel=Futures Price,
                    ylabel style={overlay},
                    xmin=-0.1, xmax=2.1,
                    xtick={0.085,1.085,2.085},
                    xticklabels={{\tt JAN22},{\tt JAN23},{\tt JAN24}},
                    ymin=203, ymax=227,
                    grid=major,
                    legend style={legend pos=south west},
                    axis background/.style={fill=gray!10}]
\addplot [color=mydarkblue,thick,smooth] table [y=Fstag1,x=T] from \heatingoilamsh;
\addplot [color=mydarkblue,thick,smooth,style=dotted] table [y=Fplain1,x=T] from \heatingoilamsh;
\addplot [color=mydarkmagenta,thick,smooth] table [y=Fbenth1,x=T] from \heatingoilamsh;
\addplot [color=mydarkred,thick,only marks,mark=triangle,mark options={yshift=(\pgfplotmarksize+\pgflinewidth)/2}] table [y=Mbid1,x=T] from \heatingoilamsh;
\addplot [color=mydarkred,thick,only marks,mark=triangle,mark options={yshift=(\pgflinewidth-\pgfplotmarksize)/2,style={rotate=180}}] table [y=Mask1,x=T] from \heatingoilamsh;
\end{axis}
\end{tikzpicture}}
\hspace*{0.25cm}
\scalebox{0.9}{%
\begin{tikzpicture}
\begin{axis}[xlabel=Maturity,
                    ylabel=Futures Price,
                    ylabel style={overlay},
                    xmin=-0.1, xmax=2.1,
                    xtick={0.085,1.085,2.085},
                    xticklabels={{\tt JAN22},{\tt JAN23},{\tt JAN24}},
                    ymin=203, ymax=227,
                    grid=major,
                    legend style={legend pos=south west},
                    axis background/.style={fill=gray!10}]
\addplot [color=mydarkblue,thick,smooth] table [y=Fstag3,x=T] from \heatingoilamsh;
\addplot [color=mydarkblue,thick,smooth,style=dotted] table [y=Fplain3,x=T] from \heatingoilamsh;
\addplot [color=mydarkmagenta,thick,smooth] table [y=Fbenth3,x=T] from \heatingoilamsh;
\addplot [color=mydarkred,thick,only marks,mark=triangle,mark options={yshift=(\pgfplotmarksize+\pgflinewidth)/2}] table [y=Mbid3,x=T] from \heatingoilamsh;
\addplot [color=mydarkred,thick,only marks,mark=triangle,mark options={yshift=(\pgflinewidth-\pgfplotmarksize)/2,style={rotate=180}}] table [y=Mask3,x=T] from \heatingoilamsh;
\end{axis}
\end{tikzpicture}}
\\\vspace*{0.5cm}
\scalebox{0.9}{%
\begin{tikzpicture}
\begin{axis}[xlabel=Maturity,
                    ylabel=Futures Price,
                    ylabel style={overlay},
                    xmin=-0.1, xmax=2.1,
                    xtick={0.085,1.085,2.085},
                    xticklabels={{\tt JAN22},{\tt JAN23},{\tt JAN24}},
                    ymin=203, ymax=227,
                    grid=major,
                    legend style={legend pos=south west},
                    axis background/.style={fill=gray!10}]
\addplot [color=mydarkblue,thick,smooth] table [y=Fstag6,x=T] from \heatingoilamsh;
\addplot [color=mydarkblue,thick,smooth,style=dotted] table [y=Fplain6,x=T] from \heatingoilamsh;
\addplot [color=mydarkmagenta,thick,smooth] table [y=Fbenth6,x=T] from \heatingoilamsh;
\addplot [color=mydarkred,thick,only marks,mark=triangle,mark options={yshift=(\pgfplotmarksize+\pgflinewidth)/2}] table [y=Mbid6,x=T] from \heatingoilamsh;
\addplot [color=mydarkred,thick,only marks,mark=triangle,mark options={yshift=(\pgflinewidth-\pgfplotmarksize)/2,style={rotate=180}}] table [y=Mask6,x=T] from \heatingoilamsh;
\end{axis}
\end{tikzpicture}}
\hspace*{0.25cm}
\scalebox{0.9}{%
\begin{tikzpicture}
\begin{axis}[xlabel=Maturity,
                    ylabel=Futures Price,
                    ylabel style={overlay},
                    xmin=-0.1, xmax=2.1,
                    xtick={0.085,1.085,2.085},
                    xticklabels={{\tt JAN22},{\tt JAN23},{\tt JAN24}},
                    ymin=203, ymax=227,
                    grid=major,
                    legend style={legend pos=south west},
                    axis background/.style={fill=gray!10}]
\addplot [color=mydarkblue,thick,smooth] table [y=Fstag12,x=T] from \heatingoilamsh;
\addplot [color=mydarkblue,thick,smooth,style=dotted] table [y=Fplain12,x=T] from \heatingoilamsh;
\addplot [color=mydarkmagenta,thick,smooth] table [y=Fbenth12,x=T] from \heatingoilamsh;
\addplot [color=mydarkred,thick,only marks,mark=triangle,mark options={yshift=(\pgfplotmarksize+\pgflinewidth)/2}] table [y=Mbid12,x=T] from \heatingoilamsh;
\addplot [color=mydarkred,thick,only marks,mark=triangle,mark options={yshift=(\pgflinewidth-\pgfplotmarksize)/2,style={rotate=180}}] table [y=Mask12,x=T] from \heatingoilamsh;
\end{axis}
\end{tikzpicture}}
\end{center}
\caption{Term structures of futures prices for Heating Oil quoted on 15 December 2021 on NYMEX market. From top-left to bottom-right the panels show the delivery periods of one, three, six months and one year. Triangles represent bid-ask futures contracts prices. Continuous blue lines is our model with penalization, dotted blue lines without it, while continuous pink lines is classical Kriging.}
\label{fig:heatingoila}
\end{figure}

\begin{table}
\begin{center}
\pgfplotstabletypeset[
    col sep=tab,
    columns={id,bid,ask,Fplain,Fstag,Fbenth},
    columns/id/.style={column name=Contract, column type={c|}, string type},
    columns/bid/.style={column name=$q^b$, fixed, fixed zerofill, precision=2, dec sep align, clear infinite},
    columns/ask/.style={column name=$q^a$, fixed, fixed zerofill, precision=2, dec sep align, clear infinite},
    columns/Fplain/.style={column name=$F$, fixed, fixed zerofill, precision=2, dec sep align, clear infinite},
    columns/Fstag/.style={column name=$F^{\rm K}$, fixed, fixed zerofill, precision=2, dec sep align, clear infinite},
    columns/Fbenth/.style={column name=$F^{\rm B}$, fixed, fixed zerofill, precision=2, dec sep align, clear infinite},
    every head row/.style={before row=\toprule,after row=\midrule},
    every last row/.style={after row=\bottomrule},
]{\heatingoilbmkt}
\end{center}
\caption{Market data for Heating Oil observed on NYMEX exchange on 11 March 2022. One-month contracts. The columns report from left to right contract identifier, bid and ask prices, model implied prices with and without seasonality management, classical Kriging implied prices.}
\label{tab:heatingoilb}
\end{table}

\begin{figure}
\begin{center}
\scalebox{0.9}{%
\begin{tikzpicture}
\begin{axis}[xlabel=Maturity,
                    ylabel=Futures Price,
                    ylabel style={overlay},
                    xmin=-0.1, xmax=2.1,
                    xtick={0.838,1.838},
                    xticklabels={{\tt JAN23},{\tt JAN24}},
                    ymin=243, ymax=347,
                    grid=major,
                    legend style={legend pos=south west},
                    axis background/.style={fill=gray!10}]
\addplot [color=mydarkblue,thick,smooth] table [y=Fstag1,x=T] from \heatingoilbmsh;
\addplot [color=mydarkblue,thick,smooth,style=dotted] table [y=Fplain1,x=T] from \heatingoilbmsh;
\addplot [color=mydarkmagenta,thick,smooth] table [y=Fbenth1,x=T] from \heatingoilbmsh;
\addplot [color=mydarkred,thick,only marks,mark=triangle,mark options={yshift=(\pgfplotmarksize+\pgflinewidth)/2}] table [y=Mbid1,x=T] from \heatingoilbmsh;
\addplot [color=mydarkred,thick,only marks,mark=triangle,mark options={yshift=(\pgflinewidth-\pgfplotmarksize)/2,style={rotate=180}}] table [y=Mask1,x=T] from \heatingoilbmsh;
\end{axis}
\end{tikzpicture}}
\hspace*{0.5cm}
\scalebox{0.9}{%
\begin{tikzpicture}
\begin{axis}[xlabel=Maturity,
                    ylabel=Futures Price,
                    ylabel style={overlay},
                    xmin=-0.1, xmax=2.1,
                    xtick={0.838,1.838},
                    xticklabels={{\tt JAN23},{\tt JAN24}},
                    ymin=243, ymax=347,
                    grid=major,
                    legend style={legend pos=south west},
                    axis background/.style={fill=gray!10}]
\addplot [color=mydarkblue,thick,smooth] table [y=Fstag3,x=T] from \heatingoilbmsh;
\addplot [color=mydarkblue,thick,smooth,style=dotted] table [y=Fplain3,x=T] from \heatingoilbmsh;
\addplot [color=mydarkmagenta,thick,smooth] table [y=Fbenth3,x=T] from \heatingoilbmsh;
\addplot [color=mydarkred,thick,only marks,mark=triangle,mark options={yshift=(\pgfplotmarksize+\pgflinewidth)/2}] table [y=Mbid3,x=T] from \heatingoilbmsh;
\addplot [color=mydarkred,thick,only marks,mark=triangle,mark options={yshift=(\pgflinewidth-\pgfplotmarksize)/2,style={rotate=180}}] table [y=Mask3,x=T] from \heatingoilbmsh;
\end{axis}
\end{tikzpicture}}
\\\vspace*{0.5cm}
\scalebox{0.9}{%
\begin{tikzpicture}
\begin{axis}[xlabel=Maturity,
                    ylabel=Futures Price,
                    ylabel style={overlay},
                    xmin=-0.1, xmax=2.1,
                    xtick={0.838,1.838},
                    xticklabels={{\tt JAN23},{\tt JAN24}},
                    ymin=243, ymax=347,
                    grid=major,
                    legend style={legend pos=south west},
                    axis background/.style={fill=gray!10}]
\addplot [color=mydarkblue,thick,smooth] table [y=Fstag6,x=T] from \heatingoilbmsh;
\addplot [color=mydarkblue,thick,smooth,style=dotted] table [y=Fplain6,x=T] from \heatingoilbmsh;
\addplot [color=mydarkmagenta,thick,smooth] table [y=Fbenth6,x=T] from \heatingoilbmsh;
\addplot [color=mydarkred,thick,only marks,mark=triangle,mark options={yshift=(\pgfplotmarksize+\pgflinewidth)/2}] table [y=Mbid6,x=T] from \heatingoilbmsh;
\addplot [color=mydarkred,thick,only marks,mark=triangle,mark options={yshift=(\pgflinewidth-\pgfplotmarksize)/2,style={rotate=180}}] table [y=Mask6,x=T] from \heatingoilbmsh;
\end{axis}
\end{tikzpicture}}
\hspace*{0.5cm}
\scalebox{0.9}{%
\begin{tikzpicture}
\begin{axis}[xlabel=Maturity,
                    ylabel=Futures Price,
                    ylabel style={overlay},
                    xmin=-0.1, xmax=2.1,
                    xtick={0.838,1.838},
                    xticklabels={{\tt JAN23},{\tt JAN24}},
                    ymin=243, ymax=347,
                    grid=major,
                    legend style={legend pos=south west},
                    axis background/.style={fill=gray!10}]
\addplot [color=mydarkblue,thick,smooth] table [y=Fstag12,x=T] from \heatingoilbmsh;
\addplot [color=mydarkblue,thick,smooth,style=dotted] table [y=Fplain12,x=T] from \heatingoilbmsh;
\addplot [color=mydarkmagenta,thick,smooth] table [y=Fbenth12,x=T] from \heatingoilbmsh;
\addplot [color=mydarkred,thick,only marks,mark=triangle,mark options={yshift=(\pgfplotmarksize+\pgflinewidth)/2}] table [y=Mbid12,x=T] from \heatingoilbmsh;
\addplot [color=mydarkred,thick,only marks,mark=triangle,mark options={yshift=(\pgflinewidth-\pgfplotmarksize)/2,style={rotate=180}}] table [y=Mask12,x=T] from \heatingoilbmsh;
\end{axis}
\end{tikzpicture}}
\end{center}
\caption{Term structures of futures prices for Heating Oil quoted on 11 March 2022 on NYMEX market. From top-left to bottom-right the panels show the delivery periods of one, three, six months and one year. Triangles represent bid-ask futures contracts prices. Continuous blue lines is our model with penalization, dotted blue lines without it, while continuous pink lines is classical Kriging.}
\label{fig:heatingoilb}
\end{figure}

\subsection{Metals}

We investigate various metals. In particular, we consider gold from CME on 15 December 2021 (table \ref{tab:gold}), silver from CME on 15 December 2021 (table \ref{tab:silver}), copper from CME on 15 December 2021 (table \ref{tab:copper}).

We notice that the market quotes only monthly futures. If we analyze data for all the three metals we can see that for shorter maturities, when quotations have a small bid-ask spread, all models show a reasonable pattern (figures \ref{fig:gold}, \ref{fig:silver} and \ref{fig:copper}). On the other hand for longer maturities, when bid-ask spreads widen, our model keeps a stable form (with and without penalty terms for seasonal patterns), while classical Kriging becomes noisy.

\begin{table}
\begin{center}
\pgfplotstabletypeset[
    col sep=tab,
    columns={id,bid,ask,Fplain,Fstag,Fbenth},
    columns/id/.style={column name=Contract, column type={c|}, string type},
    columns/bid/.style={column name=$q^b$, fixed, fixed zerofill, precision=2, dec sep align, clear infinite},
    columns/ask/.style={column name=$q^a$, fixed, fixed zerofill, precision=2, dec sep align, clear infinite},
    columns/Fplain/.style={column name=$F$, fixed, fixed zerofill, precision=2, dec sep align, clear infinite},
    columns/Fstag/.style={column name=$F^{\rm K}$, fixed, fixed zerofill, precision=2, dec sep align, clear infinite},
    columns/Fbenth/.style={column name=$F^{\rm B}$, fixed, fixed zerofill, precision=2, dec sep align, clear infinite},
    every head row/.style={before row=\toprule,after row=\midrule},
    every last row/.style={after row=\bottomrule},
]{\goldmkt}
\end{center}
\caption{Market data for Gold observed on CME exchange on 15 December 2021. One-month contracts. The columns report from left to right contract identifier, bid and ask prices, model implied prices with and without seasonality management, classical Kriging implied prices.}
\label{tab:gold}
\end{table}

\begin{figure}
\begin{center}
\scalebox{0.9}{%
\begin{tikzpicture}
\begin{axis}[xlabel=Maturity,
                    ylabel=Futures Price,
                    ylabel style={overlay},
                    xmin=-0.1, xmax=6.1,
                    xtick={0.085,1.085,2.085,3.085,4.088,5.088,6.088},
                    xticklabels={{\tt JAN22},{\tt JAN23},{\tt JAN24},{\tt JAN25},{\tt JAN26},{\tt JAN27},{\tt JAN28}},
                    ymin=1703, ymax=2253,
                    grid=major,
                    legend style={legend pos=south west},
                    axis background/.style={fill=gray!10}]
\addplot [color=mydarkblue,thick,smooth] table [y=Fstag1,x=T] from \goldmsh;
\addplot [color=mydarkblue,thick,smooth,style=dotted] table [y=Fplain1,x=T] from \goldmsh;
\addplot [color=mydarkmagenta,thick,smooth] table [y=Fbenth1,x=T] from \goldmsh;
\addplot [color=mydarkred,thick,only marks,mark=triangle,mark options={yshift=(\pgfplotmarksize+\pgflinewidth)/2}] table [y=Mbid1,x=T] from \goldmsh;
\addplot [color=mydarkred,thick,only marks,mark=triangle,mark options={yshift=(\pgflinewidth-\pgfplotmarksize)/2,style={rotate=180}}] table [y=Mask1,x=T] from \goldmsh;
\end{axis}
\end{tikzpicture}}
\hspace*{-0.25cm}
\scalebox{0.9}{%
\begin{tikzpicture}
\begin{axis}[xlabel=Maturity,
                    ylabel=Futures Price,
                    ylabel style={overlay},
                    xmin=-0.1, xmax=6.1,
                    xtick={0.085,1.085,2.085,3.085,4.088,5.088,6.088},
                    xticklabels={{\tt JAN22},{\tt JAN23},{\tt JAN24},{\tt JAN25},{\tt JAN26},{\tt JAN27},{\tt JAN28}},
                    ymin=1703, ymax=2253,
                    grid=major,
                    legend style={legend pos=south west},
                    axis background/.style={fill=gray!10}]
\addplot [color=mydarkblue,thick,smooth] table [y=Fstag3,x=T] from \goldmsh;
\addplot [color=mydarkblue,thick,smooth,style=dotted] table [y=Fplain3,x=T] from \goldmsh;
\addplot [color=mydarkmagenta,thick,smooth] table [y=Fbenth3,x=T] from \goldmsh;
\addplot [color=mydarkred,thick,only marks,mark=triangle,mark options={yshift=(\pgfplotmarksize+\pgflinewidth)/2}] table [y=Mbid3,x=T] from \goldmsh;
\addplot [color=mydarkred,thick,only marks,mark=triangle,mark options={yshift=(\pgflinewidth-\pgfplotmarksize)/2,style={rotate=180}}] table [y=Mask3,x=T] from \goldmsh;
\end{axis}
\end{tikzpicture}}
\\\vspace*{0.5cm}
\scalebox{0.9}{%
\begin{tikzpicture}
\begin{axis}[xlabel=Maturity,
                    ylabel=Futures Price,
                    ylabel style={overlay},
                    xmin=-0.1, xmax=6.1,
                    xtick={0.085,1.085,2.085,3.085,4.088,5.088,6.088},
                    xticklabels={{\tt JAN22},{\tt JAN23},{\tt JAN24},{\tt JAN25},{\tt JAN26},{\tt JAN27},{\tt JAN28}},
                    ymin=1703, ymax=2253,
                    grid=major,
                    legend style={legend pos=south west},
                    axis background/.style={fill=gray!10}]
\addplot [color=mydarkblue,thick,smooth] table [y=Fstag6,x=T] from \goldmsh;
\addplot [color=mydarkblue,thick,smooth,style=dotted] table [y=Fplain6,x=T] from \goldmsh;
\addplot [color=mydarkmagenta,thick,smooth] table [y=Fbenth6,x=T] from \goldmsh;
\addplot [color=mydarkred,thick,only marks,mark=triangle,mark options={yshift=(\pgfplotmarksize+\pgflinewidth)/2}] table [y=Mbid6,x=T] from \goldmsh;
\addplot [color=mydarkred,thick,only marks,mark=triangle,mark options={yshift=(\pgflinewidth-\pgfplotmarksize)/2,style={rotate=180}}] table [y=Mask6,x=T] from \goldmsh;
\end{axis}
\end{tikzpicture}}
\hspace*{-0.25cm}
\scalebox{0.9}{%
\begin{tikzpicture}
\begin{axis}[xlabel=Maturity,
                    ylabel=Futures Price,
                    ylabel style={overlay},
                    xmin=-0.1, xmax=6.1,
                    xtick={0.085,1.085,2.085,3.085,4.088,5.088,6.088},
                    xticklabels={{\tt JAN22},{\tt JAN23},{\tt JAN24},{\tt JAN25},{\tt JAN26},{\tt JAN27},{\tt JAN28}},
                    ymin=1703, ymax=2253,
                    grid=major,
                    legend style={legend pos=south west},
                    axis background/.style={fill=gray!10}]
\addplot [color=mydarkblue,thick,smooth] table [y=Fstag12,x=T] from \goldmsh;
\addplot [color=mydarkblue,thick,smooth,style=dotted] table [y=Fplain12,x=T] from \goldmsh;
\addplot [color=mydarkmagenta,thick,smooth] table [y=Fbenth12,x=T] from \goldmsh;
\addplot [color=mydarkred,thick,only marks,mark=triangle,mark options={yshift=(\pgfplotmarksize+\pgflinewidth)/2}] table [y=Mbid12,x=T] from \goldmsh;
\addplot [color=mydarkred,thick,only marks,mark=triangle,mark options={yshift=(\pgflinewidth-\pgfplotmarksize)/2,style={rotate=180}}] table [y=Mask12,x=T] from \goldmsh;
\end{axis}
\end{tikzpicture}}
\end{center}
\caption{Term structures of futures prices for Gold quoted on 15 December 2021 on CME market. From top-left to bottom-right the panels show the delivery periods of one, three, six months and one year. Triangles represent bid-ask futures contracts prices. Continuous blue lines is our model with penalization, dotted blue lines without it, while continuous pink lines is classical Kriging.}
\label{fig:gold}
\end{figure}

\begin{table}
\begin{center}
\pgfplotstabletypeset[
    col sep=tab,
    columns={id,bid,ask,Fplain,Fstag,Fbenth},
    columns/id/.style={column name=Contract, column type={c|}, string type},
    columns/bid/.style={column name=$q^b$, fixed, fixed zerofill, precision=2, dec sep align, clear infinite},
    columns/ask/.style={column name=$q^a$, fixed, fixed zerofill, precision=2, dec sep align, clear infinite},
    columns/Fplain/.style={column name=$F$, fixed, fixed zerofill, precision=2, dec sep align, clear infinite},
    columns/Fstag/.style={column name=$F^{\rm K}$, fixed, fixed zerofill, precision=2, dec sep align, clear infinite},
    columns/Fbenth/.style={column name=$F^{\rm B}$, fixed, fixed zerofill, precision=2, dec sep align, clear infinite},
    every head row/.style={before row=\toprule,after row=\midrule},
    every last row/.style={after row=\bottomrule},
]{\silvermkt}
\end{center}
\caption{Market data for Silver observed on CME exchange on 15 December 2021. One-month contracts. The columns report from left to right contract identifier, bid and ask prices, model implied prices with and without seasonality management, classical Kriging implied prices.}
\label{tab:silver}
\end{table}

\begin{figure}
\begin{center}
\scalebox{0.9}{%
\begin{tikzpicture}
\begin{axis}[xlabel=Maturity,
                    ylabel=Futures Price,
                    ylabel style={overlay},
                    xmin=-0.1, xmax=5.1,
                    xtick={0.085,1.085,2.085,3.085,4.088,5.088},
                    xticklabels={{\tt JAN22},{\tt JAN23},{\tt JAN24},{\tt JAN25},{\tt JAN26},{\tt JAN27}},
                    ymin=17, ymax=33,
                    grid=major,
                    legend style={legend pos=south west},
                    axis background/.style={fill=gray!10}]
\addplot [color=mydarkblue,thick,smooth] table [y=Fstag1,x=T] from \silvermsh;
\addplot [color=mydarkblue,thick,smooth,style=dotted] table [y=Fplain1,x=T] from \silvermsh;
\addplot [color=mydarkmagenta,thick,smooth] table [y=Fbenth1,x=T] from \silvermsh;
\addplot [color=mydarkred,thick,only marks,mark=triangle,mark options={yshift=(\pgfplotmarksize+\pgflinewidth)/2}] table [y=Mbid1,x=T] from \silvermsh;
\addplot [color=mydarkred,thick,only marks,mark=triangle,mark options={yshift=(\pgflinewidth-\pgfplotmarksize)/2,style={rotate=180}}] table [y=Mask1,x=T] from \silvermsh;
\end{axis}
\end{tikzpicture}}
\hspace*{0.5cm}
\scalebox{0.9}{%
\begin{tikzpicture}
\begin{axis}[xlabel=Maturity,
                    ylabel=Futures Price,
                    ylabel style={overlay},
                    xmin=-0.1, xmax=5.1,
                    xtick={0.085,1.085,2.085,3.085,4.088,5.088},
                    xticklabels={{\tt JAN22},{\tt JAN23},{\tt JAN24},{\tt JAN25},{\tt JAN26},{\tt JAN27}},
                    ymin=17, ymax=33,
                    grid=major,
                    legend style={legend pos=south west},
                    axis background/.style={fill=gray!10}]
\addplot [color=mydarkblue,thick,smooth] table [y=Fstag3,x=T] from \silvermsh;
\addplot [color=mydarkblue,thick,smooth,style=dotted] table [y=Fplain3,x=T] from \silvermsh;
\addplot [color=mydarkmagenta,thick,smooth] table [y=Fbenth3,x=T] from \silvermsh;
\addplot [color=mydarkred,thick,only marks,mark=triangle,mark options={yshift=(\pgfplotmarksize+\pgflinewidth)/2}] table [y=Mbid3,x=T] from \silvermsh;
\addplot [color=mydarkred,thick,only marks,mark=triangle,mark options={yshift=(\pgflinewidth-\pgfplotmarksize)/2,style={rotate=180}}] table [y=Mask3,x=T] from \silvermsh;
\end{axis}
\end{tikzpicture}}
\\\vspace*{0.5cm}
\scalebox{0.9}{%
\begin{tikzpicture}
\begin{axis}[xlabel=Maturity,
                    ylabel=Futures Price,
                    ylabel style={overlay},
                    xmin=-0.1, xmax=5.1,
                    xtick={0.085,1.085,2.085,3.085,4.088,5.088},
                    xticklabels={{\tt JAN22},{\tt JAN23},{\tt JAN24},{\tt JAN25},{\tt JAN26},{\tt JAN27}},
                    ymin=17, ymax=33,
                    grid=major,
                    legend style={legend pos=south west},
                    axis background/.style={fill=gray!10}]
\addplot [color=mydarkblue,thick,smooth] table [y=Fstag6,x=T] from \silvermsh;
\addplot [color=mydarkblue,thick,smooth,style=dotted] table [y=Fplain6,x=T] from \silvermsh;
\addplot [color=mydarkmagenta,thick,smooth] table [y=Fbenth6,x=T] from \silvermsh;
\addplot [color=mydarkred,thick,only marks,mark=triangle,mark options={yshift=(\pgfplotmarksize+\pgflinewidth)/2}] table [y=Mbid6,x=T] from \silvermsh;
\addplot [color=mydarkred,thick,only marks,mark=triangle,mark options={yshift=(\pgflinewidth-\pgfplotmarksize)/2,style={rotate=180}}] table [y=Mask6,x=T] from \silvermsh;
\end{axis}
\end{tikzpicture}}
\hspace*{0.5cm}
\scalebox{0.9}{%
\begin{tikzpicture}
\begin{axis}[xlabel=Maturity,
                    ylabel=Futures Price,
                    ylabel style={overlay},
                    xmin=-0.1, xmax=5.1,
                    xtick={0.085,1.085,2.085,3.085,4.088,5.088},
                    xticklabels={{\tt JAN22},{\tt JAN23},{\tt JAN24},{\tt JAN25},{\tt JAN26},{\tt JAN27}},
                    ymin=17, ymax=33,
                    grid=major,
                    legend style={legend pos=south west},
                    axis background/.style={fill=gray!10}]
\addplot [color=mydarkblue,thick,smooth] table [y=Fstag12,x=T] from \silvermsh;
\addplot [color=mydarkblue,thick,smooth,style=dotted] table [y=Fplain12,x=T] from \silvermsh;
\addplot [color=mydarkmagenta,thick,smooth] table [y=Fbenth12,x=T] from \silvermsh;
\addplot [color=mydarkred,thick,only marks,mark=triangle,mark options={yshift=(\pgfplotmarksize+\pgflinewidth)/2}] table [y=Mbid12,x=T] from \silvermsh;
\addplot [color=mydarkred,thick,only marks,mark=triangle,mark options={yshift=(\pgflinewidth-\pgfplotmarksize)/2,style={rotate=180}}] table [y=Mask12,x=T] from \silvermsh;
\end{axis}
\end{tikzpicture}}
\end{center}
\caption{Term structures of futures prices for Silver quoted on 15 December 2021 on CME market. From top-left to bottom-right the panels show the delivery periods of one, three, six months and one year. Triangles represent bid-ask futures contracts prices. Continuous blue lines is our model with penalization, dotted blue lines without it, while continuous pink lines is classical Kriging.}
\label{fig:silver}
\end{figure}

\begin{table}
\begin{center}
\pgfplotstabletypeset[
    col sep=tab,
    columns={id,bid,ask,Fplain,Fstag,Fbenth},
    columns/id/.style={column name=Contract, column type={c|}, string type},
    columns/bid/.style={column name=$q^b$, fixed, fixed zerofill, precision=2, dec sep align, clear infinite},
    columns/ask/.style={column name=$q^a$, fixed, fixed zerofill, precision=2, dec sep align, clear infinite},
    columns/Fplain/.style={column name=$F$, fixed, fixed zerofill, precision=2, dec sep align, clear infinite},
    columns/Fstag/.style={column name=$F^{\rm K}$, fixed, fixed zerofill, precision=2, dec sep align, clear infinite},
    columns/Fbenth/.style={column name=$F^{\rm B}$, fixed, fixed zerofill, precision=2, dec sep align, clear infinite},
    every head row/.style={before row=\toprule,after row=\midrule},
    every last row/.style={after row=\bottomrule},
]{\coppermkt}
\end{center}
\caption{Market data for Copper observed on CME exchange on 15 December 2021. One-month contracts. The columns report from left to right contract identifier, bid and ask prices, model implied prices with and without seasonality management, classical Kriging implied prices.}
\label{tab:copper}
\end{table}

\begin{figure}
\begin{center}
\scalebox{0.9}{%
\begin{tikzpicture}
\begin{axis}[xlabel=Maturity,
                    ylabel=Futures Price,
                    ylabel style={overlay},
                    xmin=-0.1, xmax=3.1,
                    xtick={0.085,1.085,2.085,3.085},
                    xticklabels={{\tt JAN22},{\tt JAN23},{\tt JAN24},{\tt JAN25}},
                    ymin=387, ymax=423,
                    grid=major,
                    legend style={legend pos=south west},
                    axis background/.style={fill=gray!10}]
\addplot [color=mydarkblue,thick,smooth] table [y=Fstag1,x=T] from \coppermsh;
\addplot [color=mydarkblue,thick,smooth,style=dotted] table [y=Fplain1,x=T] from \coppermsh;
\addplot [color=mydarkmagenta,thick,smooth] table [y=Fbenth1,x=T] from \coppermsh;
\addplot [color=mydarkred,thick,only marks,mark=triangle,mark options={yshift=(\pgfplotmarksize+\pgflinewidth)/2}] table [y=Mbid1,x=T] from \coppermsh;
\addplot [color=mydarkred,thick,only marks,mark=triangle,mark options={yshift=(\pgflinewidth-\pgfplotmarksize)/2,style={rotate=180}}] table [y=Mask1,x=T] from \coppermsh;
\end{axis}
\end{tikzpicture}}
\hspace*{0.25cm}
\scalebox{0.9}{%
\begin{tikzpicture}
\begin{axis}[xlabel=Maturity,
                    ylabel=Futures Price,
                    ylabel style={overlay},
                    xmin=-0.1, xmax=3.1,
                    xtick={0.085,1.085,2.085,3.085},
                    xticklabels={{\tt JAN22},{\tt JAN23},{\tt JAN24},{\tt JAN25}},
                    ymin=387, ymax=423,
                    grid=major,
                    legend style={legend pos=south west},
                    axis background/.style={fill=gray!10}]
\addplot [color=mydarkblue,thick,smooth] table [y=Fstag3,x=T] from \coppermsh;
\addplot [color=mydarkblue,thick,smooth,style=dotted] table [y=Fplain3,x=T] from \coppermsh;
\addplot [color=mydarkmagenta,thick,smooth] table [y=Fbenth3,x=T] from \coppermsh;
\addplot [color=mydarkred,thick,only marks,mark=triangle,mark options={yshift=(\pgfplotmarksize+\pgflinewidth)/2}] table [y=Mbid3,x=T] from \coppermsh;
\addplot [color=mydarkred,thick,only marks,mark=triangle,mark options={yshift=(\pgflinewidth-\pgfplotmarksize)/2,style={rotate=180}}] table [y=Mask3,x=T] from \coppermsh;
\end{axis}
\end{tikzpicture}}
\\\vspace*{0.5cm}
\scalebox{0.9}{%
\begin{tikzpicture}
\begin{axis}[xlabel=Maturity,
                    ylabel=Futures Price,
                    ylabel style={overlay},
                    xmin=-0.1, xmax=3.1,
                    xtick={0.085,1.085,2.085,3.085},
                    xticklabels={{\tt JAN22},{\tt JAN23},{\tt JAN24},{\tt JAN25}},
                    ymin=387, ymax=423,
                    grid=major,
                    legend style={legend pos=south west},
                    axis background/.style={fill=gray!10}]
\addplot [color=mydarkblue,thick,smooth] table [y=Fstag6,x=T] from \coppermsh;
\addplot [color=mydarkblue,thick,smooth,style=dotted] table [y=Fplain6,x=T] from \coppermsh;
\addplot [color=mydarkmagenta,thick,smooth] table [y=Fbenth6,x=T] from \coppermsh;
\addplot [color=mydarkred,thick,only marks,mark=triangle,mark options={yshift=(\pgfplotmarksize+\pgflinewidth)/2}] table [y=Mbid6,x=T] from \coppermsh;
\addplot [color=mydarkred,thick,only marks,mark=triangle,mark options={yshift=(\pgflinewidth-\pgfplotmarksize)/2,style={rotate=180}}] table [y=Mask6,x=T] from \coppermsh;
\end{axis}
\end{tikzpicture}}
\hspace*{0.25cm}
\scalebox{0.9}{%
\begin{tikzpicture}
\begin{axis}[xlabel=Maturity,
                    ylabel=Futures Price,
                    ylabel style={overlay},
                    xmin=-0.1, xmax=3.1,
                    xtick={0.085,1.085,2.085,3.085},
                    xticklabels={{\tt JAN22},{\tt JAN23},{\tt JAN24},{\tt JAN25}},
                    ymin=387, ymax=423,
                    grid=major,
                    legend style={legend pos=south west},
                    axis background/.style={fill=gray!10}]
\addplot [color=mydarkblue,thick,smooth] table [y=Fstag12,x=T] from \coppermsh;
\addplot [color=mydarkblue,thick,smooth,style=dotted] table [y=Fplain12,x=T] from \coppermsh;
\addplot [color=mydarkmagenta,thick,smooth] table [y=Fbenth12,x=T] from \coppermsh;
\addplot [color=mydarkred,thick,only marks,mark=triangle,mark options={yshift=(\pgfplotmarksize+\pgflinewidth)/2}] table [y=Mbid12,x=T] from \coppermsh;
\addplot [color=mydarkred,thick,only marks,mark=triangle,mark options={yshift=(\pgflinewidth-\pgfplotmarksize)/2,style={rotate=180}}] table [y=Mask12,x=T] from \coppermsh;
\end{axis}
\end{tikzpicture}}
\end{center}
\caption{Term structures of futures prices for Copper quoted on 15 December 2021 on CME market. From top-left to bottom-right the panels show the delivery periods of one, three, six months and one year. Triangles represent bid-ask futures contracts prices. Continuous blue lines is our model with penalization, dotted blue lines without it, while continuous pink lines is classical Kriging.}
\label{fig:copper}
\end{figure}

\subsection{Agricultural goods}

We investigate various agricultural goods. In particular, we consider sugar {no.11} from ICE on 15 December 2021 (table \ref{tab:sugar}), coffee C from ICE on 15 December 2021 (table \ref{tab:coffee}), cocoa from ICE on 15 December 2021 (table \ref{tab:cocoa}), soybean from CBOE on 15 December 2021 (table \ref{tab:soybean}), wheat from CBOE on 15 December 2021 (table \ref{tab:wheat}).

We notice that the market quotes only monthly futures. If we analyze data for all the five goods we can see various term structure shapes: small seasonal patterns for sugar (figures \ref{fig:sugar}), a decreasing trend for coffee (figures \ref{fig:coffee}), more or less irregular shapes for cocoa, soybean and wheat (figures \ref{fig:cocoa}, \ref{fig:soybean} and \ref{fig:wheat}). Our model adapts to all the behaviours and keeps a stable extrapolation (with or without penalty terms for seasonal patterns), while classical Kriging becomes noisy when bid-ask spreads widen and in the extrapolation region.

\begin{table}
\begin{center}
\pgfplotstabletypeset[
    col sep=tab,
    columns={id,bid,ask,Fplain,Fstag,Fbenth},
    columns/id/.style={column name=Contract, column type={c|}, string type},
    columns/bid/.style={column name=$q^b$, fixed, fixed zerofill, precision=2, dec sep align, clear infinite},
    columns/ask/.style={column name=$q^a$, fixed, fixed zerofill, precision=2, dec sep align, clear infinite},
    columns/Fplain/.style={column name=$F$, fixed, fixed zerofill, precision=2, dec sep align, clear infinite},
    columns/Fstag/.style={column name=$F^{\rm K}$, fixed, fixed zerofill, precision=2, dec sep align, clear infinite},
    columns/Fbenth/.style={column name=$F^{\rm B}$, fixed, fixed zerofill, precision=2, dec sep align, clear infinite},
    every head row/.style={before row=\toprule,after row=\midrule},
    every last row/.style={after row=\bottomrule},
]{\sugarmkt}
\end{center}
\caption{Market data for Sugar {No.11} observed on ICE exchange on 15 December 2021. One-month contracts. The columns report from left to right contract identifier, bid and ask prices, model implied prices with and without seasonality management, classical Kriging implied prices.}
\label{tab:sugar}
\end{table}

\begin{figure}
\begin{center}
\scalebox{0.9}{%
\begin{tikzpicture}
\begin{axis}[xlabel=Maturity,
                    ylabel=Futures Price,
                    ylabel style={overlay},
                    xmin=-0.1, xmax=3.1,
                    xtick={0.085,1.085,2.085,3.085},
                    xticklabels={{\tt JAN22},{\tt JAN23},{\tt JAN24},{\tt JAN25}},
                    ymin=13, ymax=23,
                    grid=major,
                    legend style={legend pos=south west},
                    axis background/.style={fill=gray!10}]
\addplot [color=mydarkblue,thick,smooth] table [y=Fstag1,x=T] from \sugarmsh;
\addplot [color=mydarkblue,thick,smooth,style=dotted] table [y=Fplain1,x=T] from \sugarmsh;
\addplot [color=mydarkmagenta,thick,smooth] table [y=Fbenth1,x=T] from \sugarmsh;
\addplot [color=mydarkred,thick,only marks,mark=triangle,mark options={yshift=(\pgfplotmarksize+\pgflinewidth)/2}] table [y=Mbid1,x=T] from \sugarmsh;
\addplot [color=mydarkred,thick,only marks,mark=triangle,mark options={yshift=(\pgflinewidth-\pgfplotmarksize)/2,style={rotate=180}}] table [y=Mask1,x=T] from \sugarmsh;
\end{axis}
\end{tikzpicture}}
\hspace*{0.5cm}
\scalebox{0.9}{%
\begin{tikzpicture}
\begin{axis}[xlabel=Maturity,
                    ylabel=Futures Price,
                    ylabel style={overlay},
                    xmin=-0.1, xmax=3.1,
                    xtick={0.085,1.085,2.085,3.085},
                    xticklabels={{\tt JAN22},{\tt JAN23},{\tt JAN24},{\tt JAN25}},
                    ymin=13, ymax=23,
                    grid=major,
                    legend style={legend pos=south west},
                    axis background/.style={fill=gray!10}]
\addplot [color=mydarkblue,thick,smooth] table [y=Fstag3,x=T] from \sugarmsh;
\addplot [color=mydarkblue,thick,smooth,style=dotted] table [y=Fplain3,x=T] from \sugarmsh;
\addplot [color=mydarkmagenta,thick,smooth] table [y=Fbenth3,x=T] from \sugarmsh;
\addplot [color=mydarkred,thick,only marks,mark=triangle,mark options={yshift=(\pgfplotmarksize+\pgflinewidth)/2}] table [y=Mbid3,x=T] from \sugarmsh;
\addplot [color=mydarkred,thick,only marks,mark=triangle,mark options={yshift=(\pgflinewidth-\pgfplotmarksize)/2,style={rotate=180}}] table [y=Mask3,x=T] from \sugarmsh;
\end{axis}
\end{tikzpicture}}
\\\vspace*{0.5cm}
\scalebox{0.9}{%
\begin{tikzpicture}
\begin{axis}[xlabel=Maturity,
                    ylabel=Futures Price,
                    ylabel style={overlay},
                    xmin=-0.1, xmax=3.1,
                    xtick={0.085,1.085,2.085,3.085},
                    xticklabels={{\tt JAN22},{\tt JAN23},{\tt JAN24},{\tt JAN25}},
                    ymin=13, ymax=23,
                    grid=major,
                    legend style={legend pos=south west},
                    axis background/.style={fill=gray!10}]
\addplot [color=mydarkblue,thick,smooth] table [y=Fstag6,x=T] from \sugarmsh;
\addplot [color=mydarkblue,thick,smooth,style=dotted] table [y=Fplain6,x=T] from \sugarmsh;
\addplot [color=mydarkmagenta,thick,smooth] table [y=Fbenth6,x=T] from \sugarmsh;
\addplot [color=mydarkred,thick,only marks,mark=triangle,mark options={yshift=(\pgfplotmarksize+\pgflinewidth)/2}] table [y=Mbid6,x=T] from \sugarmsh;
\addplot [color=mydarkred,thick,only marks,mark=triangle,mark options={yshift=(\pgflinewidth-\pgfplotmarksize)/2,style={rotate=180}}] table [y=Mask6,x=T] from \sugarmsh;
\end{axis}
\end{tikzpicture}}
\hspace*{0.5cm}
\scalebox{0.9}{%
\begin{tikzpicture}
\begin{axis}[xlabel=Maturity,
                    ylabel=Futures Price,
                    ylabel style={overlay},
                    xmin=-0.1, xmax=3.1,
                    xtick={0.085,1.085,2.085,3.085},
                    xticklabels={{\tt JAN22},{\tt JAN23},{\tt JAN24},{\tt JAN25}},
                    ymin=13, ymax=23,
                    grid=major,
                    legend style={legend pos=south west},
                    axis background/.style={fill=gray!10}]
\addplot [color=mydarkblue,thick,smooth] table [y=Fstag12,x=T] from \sugarmsh;
\addplot [color=mydarkblue,thick,smooth,style=dotted] table [y=Fplain12,x=T] from \sugarmsh;
\addplot [color=mydarkmagenta,thick,smooth] table [y=Fbenth12,x=T] from \sugarmsh;
\addplot [color=mydarkred,thick,only marks,mark=triangle,mark options={yshift=(\pgfplotmarksize+\pgflinewidth)/2}] table [y=Mbid12,x=T] from \sugarmsh;
\addplot [color=mydarkred,thick,only marks,mark=triangle,mark options={yshift=(\pgflinewidth-\pgfplotmarksize)/2,style={rotate=180}}] table [y=Mask12,x=T] from \sugarmsh;
\end{axis}
\end{tikzpicture}}
\end{center}
\caption{Term structures of futures prices for Sugar {No.11} quoted on 15 December 2021 on ICE market. From top-left to bottom-right the panels show the delivery periods of one, three, six months and one year. Triangles represent bid-ask futures contracts prices. Continuous blue lines is our model with penalization, dotted blue lines without it, while continuous pink lines is classical Kriging.}
\label{fig:sugar}
\end{figure}

\begin{table}
\begin{center}
\pgfplotstabletypeset[
    col sep=tab,
    columns={id,bid,ask,Fplain,Fstag,Fbenth},
    columns/id/.style={column name=Contract, column type={c|}, string type},
    columns/bid/.style={column name=$q^b$, fixed, fixed zerofill, precision=2, dec sep align, clear infinite},
    columns/ask/.style={column name=$q^a$, fixed, fixed zerofill, precision=2, dec sep align, clear infinite},
    columns/Fplain/.style={column name=$F$, fixed, fixed zerofill, precision=2, dec sep align, clear infinite},
    columns/Fstag/.style={column name=$F^{\rm K}$, fixed, fixed zerofill, precision=2, dec sep align, clear infinite},
    columns/Fbenth/.style={column name=$F^{\rm B}$, fixed, fixed zerofill, precision=2, dec sep align, clear infinite},
    every head row/.style={before row=\toprule,after row=\midrule},
    every last row/.style={after row=\bottomrule},
]{\coffeemkt}
\end{center}
\caption{Market data for Coffee C observed on ICE exchange on 15 December 2021. One-month contracts. The columns report from left to right contract identifier, bid and ask prices, model implied prices with and without seasonality management, classical Kriging implied prices.}
\label{tab:coffee}
\end{table}

\begin{figure}
\begin{center}
\scalebox{0.9}{%
\begin{tikzpicture}
\begin{axis}[xlabel=Maturity,
                    ylabel=Futures Price,
                    ylabel style={overlay},
                    xmin=-0.1, xmax=3.1,
                    xtick={0.085,1.085,2.085,3.085},
                    xticklabels={{\tt JAN22},{\tt JAN23},{\tt JAN24},{\tt JAN25}},
                    ymin=223, ymax=243,
                    grid=major,
                    legend style={legend pos=south west},
                    axis background/.style={fill=gray!10}]
\addplot [color=mydarkblue,thick,smooth] table [y=Fstag1,x=T] from \coffeemsh;
\addplot [color=mydarkblue,thick,smooth,style=dotted] table [y=Fplain1,x=T] from \coffeemsh;
\addplot [color=mydarkmagenta,thick,smooth] table [y=Fbenth1,x=T] from \coffeemsh;
\addplot [color=mydarkred,thick,only marks,mark=triangle,mark options={yshift=(\pgfplotmarksize+\pgflinewidth)/2}] table [y=Mbid1,x=T] from \coffeemsh;
\addplot [color=mydarkred,thick,only marks,mark=triangle,mark options={yshift=(\pgflinewidth-\pgfplotmarksize)/2,style={rotate=180}}] table [y=Mask1,x=T] from \coffeemsh;
\end{axis}
\end{tikzpicture}}
\hspace*{0.25cm}
\scalebox{0.9}{%
\begin{tikzpicture}
\begin{axis}[xlabel=Maturity,
                    ylabel=Futures Price,
                    ylabel style={overlay},
                    xmin=-0.1, xmax=3.1,
                    xtick={0.085,1.085,2.085,3.085},
                    xticklabels={{\tt JAN22},{\tt JAN23},{\tt JAN24},{\tt JAN25}},
                    ymin=223, ymax=243,
                    grid=major,
                    legend style={legend pos=south west},
                    axis background/.style={fill=gray!10}]
\addplot [color=mydarkblue,thick,smooth] table [y=Fstag3,x=T] from \coffeemsh;
\addplot [color=mydarkblue,thick,smooth,style=dotted] table [y=Fplain3,x=T] from \coffeemsh;
\addplot [color=mydarkmagenta,thick,smooth] table [y=Fbenth3,x=T] from \coffeemsh;
\addplot [color=mydarkred,thick,only marks,mark=triangle,mark options={yshift=(\pgfplotmarksize+\pgflinewidth)/2}] table [y=Mbid3,x=T] from \coffeemsh;
\addplot [color=mydarkred,thick,only marks,mark=triangle,mark options={yshift=(\pgflinewidth-\pgfplotmarksize)/2,style={rotate=180}}] table [y=Mask3,x=T] from \coffeemsh;
\end{axis}
\end{tikzpicture}}
\\\vspace*{0.5cm}
\scalebox{0.9}{%
\begin{tikzpicture}
\begin{axis}[xlabel=Maturity,
                    ylabel=Futures Price,
                    ylabel style={overlay},
                    xmin=-0.1, xmax=3.1,
                    xtick={0.085,1.085,2.085,3.085},
                    xticklabels={{\tt JAN22},{\tt JAN23},{\tt JAN24},{\tt JAN25}},
                    ymin=223, ymax=243,
                    grid=major,
                    legend style={legend pos=south west},
                    axis background/.style={fill=gray!10}]
\addplot [color=mydarkblue,thick,smooth] table [y=Fstag6,x=T] from \coffeemsh;
\addplot [color=mydarkblue,thick,smooth,style=dotted] table [y=Fplain6,x=T] from \coffeemsh;
\addplot [color=mydarkmagenta,thick,smooth] table [y=Fbenth6,x=T] from \coffeemsh;
\addplot [color=mydarkred,thick,only marks,mark=triangle,mark options={yshift=(\pgfplotmarksize+\pgflinewidth)/2}] table [y=Mbid6,x=T] from \coffeemsh;
\addplot [color=mydarkred,thick,only marks,mark=triangle,mark options={yshift=(\pgflinewidth-\pgfplotmarksize)/2,style={rotate=180}}] table [y=Mask6,x=T] from \coffeemsh;
\end{axis}
\end{tikzpicture}}
\hspace*{0.25cm}
\scalebox{0.9}{%
\begin{tikzpicture}
\begin{axis}[xlabel=Maturity,
                    ylabel=Futures Price,
                    ylabel style={overlay},
                    xmin=-0.1, xmax=3.1,
                    xtick={0.085,1.085,2.085,3.085},
                    xticklabels={{\tt JAN22},{\tt JAN23},{\tt JAN24},{\tt JAN25}},
                    ymin=223, ymax=243,
                    grid=major,
                    legend style={legend pos=south west},
                    axis background/.style={fill=gray!10}]
\addplot [color=mydarkblue,thick,smooth] table [y=Fstag12,x=T] from \coffeemsh;
\addplot [color=mydarkblue,thick,smooth,style=dotted] table [y=Fplain12,x=T] from \coffeemsh;
\addplot [color=mydarkmagenta,thick,smooth] table [y=Fbenth12,x=T] from \coffeemsh;
\addplot [color=mydarkred,thick,only marks,mark=triangle,mark options={yshift=(\pgfplotmarksize+\pgflinewidth)/2}] table [y=Mbid12,x=T] from \coffeemsh;
\addplot [color=mydarkred,thick,only marks,mark=triangle,mark options={yshift=(\pgflinewidth-\pgfplotmarksize)/2,style={rotate=180}}] table [y=Mask12,x=T] from \coffeemsh;
\end{axis}
\end{tikzpicture}}
\end{center}
\caption{Term structures of futures prices for Coffee C quoted on 15 December 2021 on ICE market. From top-left to bottom-right the panels show the delivery periods of one, three, six months and one year. Triangles represent bid-ask futures contracts prices. Continuous blue lines is our model with penalization, dotted blue lines without it, while continuous pink lines is classical Kriging.}
\label{fig:coffee}
\end{figure}

\begin{table}
\begin{center}
\pgfplotstabletypeset[
    col sep=tab,
    columns={id,bid,ask,Fplain,Fstag,Fbenth},
    columns/id/.style={column name=Contract, column type={c|}, string type},
    columns/bid/.style={column name=$q^b$, fixed, fixed zerofill, precision=2, dec sep align, clear infinite},
    columns/ask/.style={column name=$q^a$, fixed, fixed zerofill, precision=2, dec sep align, clear infinite},
    columns/Fplain/.style={column name=$F$, fixed, fixed zerofill, precision=2, dec sep align, clear infinite},
    columns/Fstag/.style={column name=$F^{\rm K}$, fixed, fixed zerofill, precision=2, dec sep align, clear infinite},
    columns/Fbenth/.style={column name=$F^{\rm B}$, fixed, fixed zerofill, precision=2, dec sep align, clear infinite},
    every head row/.style={before row=\toprule,after row=\midrule},
    every last row/.style={after row=\bottomrule},
]{\cocoamkt}
\end{center}
\caption{Market data for Cocoa observed on ICE exchange on 15 December 2021. One-month contracts. The columns report from left to right contract identifier, bid and ask prices, model implied prices with and without seasonality management, classical Kriging implied prices.}
\label{tab:cocoa}
\end{table}

\begin{figure}
\begin{center}
\scalebox{0.9}{%
\begin{tikzpicture}
\begin{axis}[xlabel=Maturity,
                    ylabel=Futures Price,
                    ylabel style={overlay},
                    xmin=-0.1, xmax=2.1,
                    xtick={0.085,1.085,2.085},
                    xticklabels={{\tt JAN22},{\tt JAN23},{\tt JAN24}},
                    ymin=2503, ymax=2653,
                    grid=major,
                    legend style={legend pos=south west},
                    axis background/.style={fill=gray!10}]
\addplot [color=mydarkblue,thick,smooth] table [y=Fstag1,x=T] from \cocoamsh;
\addplot [color=mydarkblue,thick,smooth,style=dotted] table [y=Fplain1,x=T] from \cocoamsh;
\addplot [color=mydarkmagenta,thick,smooth] table [y=Fbenth1,x=T] from \cocoamsh;
\addplot [color=mydarkred,thick,only marks,mark=triangle,mark options={yshift=(\pgfplotmarksize+\pgflinewidth)/2}] table [y=Mbid1,x=T] from \cocoamsh;
\addplot [color=mydarkred,thick,only marks,mark=triangle,mark options={yshift=(\pgflinewidth-\pgfplotmarksize)/2,style={rotate=180}}] table [y=Mask1,x=T] from \cocoamsh;
\end{axis}
\end{tikzpicture}}
\hspace*{-0.25cm}
\scalebox{0.9}{%
\begin{tikzpicture}
\begin{axis}[xlabel=Maturity,
                    ylabel=Futures Price,
                    ylabel style={overlay},
                    xmin=-0.1, xmax=2.1,
                    xtick={0.085,1.085,2.085},
                    xticklabels={{\tt JAN22},{\tt JAN23},{\tt JAN24}},
                    ymin=2503, ymax=2653,
                    grid=major,
                    legend style={legend pos=south west},
                    axis background/.style={fill=gray!10}]
\addplot [color=mydarkblue,thick,smooth] table [y=Fstag3,x=T] from \cocoamsh;
\addplot [color=mydarkblue,thick,smooth,style=dotted] table [y=Fplain3,x=T] from \cocoamsh;
\addplot [color=mydarkmagenta,thick,smooth] table [y=Fbenth3,x=T] from \cocoamsh;
\addplot [color=mydarkred,thick,only marks,mark=triangle,mark options={yshift=(\pgfplotmarksize+\pgflinewidth)/2}] table [y=Mbid3,x=T] from \cocoamsh;
\addplot [color=mydarkred,thick,only marks,mark=triangle,mark options={yshift=(\pgflinewidth-\pgfplotmarksize)/2,style={rotate=180}}] table [y=Mask3,x=T] from \cocoamsh;
\end{axis}
\end{tikzpicture}}
\\\vspace*{0.5cm}
\scalebox{0.9}{%
\begin{tikzpicture}
\begin{axis}[xlabel=Maturity,
                    ylabel=Futures Price,
                    ylabel style={overlay},
                    xmin=-0.1, xmax=2.1,
                    xtick={0.085,1.085,2.085},
                    xticklabels={{\tt JAN22},{\tt JAN23},{\tt JAN24}},
                    ymin=2503, ymax=2653,
                    grid=major,
                    legend style={legend pos=south west},
                    axis background/.style={fill=gray!10}]
\addplot [color=mydarkblue,thick,smooth] table [y=Fstag6,x=T] from \cocoamsh;
\addplot [color=mydarkblue,thick,smooth,style=dotted] table [y=Fplain6,x=T] from \cocoamsh;
\addplot [color=mydarkmagenta,thick,smooth] table [y=Fbenth6,x=T] from \cocoamsh;
\addplot [color=mydarkred,thick,only marks,mark=triangle,mark options={yshift=(\pgfplotmarksize+\pgflinewidth)/2}] table [y=Mbid6,x=T] from \cocoamsh;
\addplot [color=mydarkred,thick,only marks,mark=triangle,mark options={yshift=(\pgflinewidth-\pgfplotmarksize)/2,style={rotate=180}}] table [y=Mask6,x=T] from \cocoamsh;
\end{axis}
\end{tikzpicture}}
\hspace*{-0.25cm}
\scalebox{0.9}{%
\begin{tikzpicture}
\begin{axis}[xlabel=Maturity,
                    ylabel=Futures Price,
                    ylabel style={overlay},
                    xmin=-0.1, xmax=2.1,
                    xtick={0.085,1.085,2.085},
                    xticklabels={{\tt JAN22},{\tt JAN23},{\tt JAN24}},
                    ymin=2503, ymax=2653,
                    grid=major,
                    legend style={legend pos=south west},
                    axis background/.style={fill=gray!10}]
\addplot [color=mydarkblue,thick,smooth] table [y=Fstag12,x=T] from \cocoamsh;
\addplot [color=mydarkblue,thick,smooth,style=dotted] table [y=Fplain12,x=T] from \cocoamsh;
\addplot [color=mydarkmagenta,thick,smooth] table [y=Fbenth12,x=T] from \cocoamsh;
\addplot [color=mydarkred,thick,only marks,mark=triangle,mark options={yshift=(\pgfplotmarksize+\pgflinewidth)/2}] table [y=Mbid12,x=T] from \cocoamsh;
\addplot [color=mydarkred,thick,only marks,mark=triangle,mark options={yshift=(\pgflinewidth-\pgfplotmarksize)/2,style={rotate=180}}] table [y=Mask12,x=T] from \cocoamsh;
\end{axis}
\end{tikzpicture}}
\end{center}
\caption{Term structures of futures prices for Cocoa quoted on 15 December 2021 on ICE market. From top-left to bottom-right the panels show the delivery periods of one, three, six months and one year. Triangles represent bid-ask futures contracts prices. Continuous blue lines is our model with penalization, dotted blue lines without it, while continuous pink lines is classical Kriging.}
\label{fig:cocoa}
\end{figure}

\begin{table}
\begin{center}
\pgfplotstabletypeset[
    col sep=tab,
    columns={id,bid,ask,Fplain,Fstag,Fbenth},
    columns/id/.style={column name=Contract, column type={c|}, string type},
    columns/bid/.style={column name=$q^b$, fixed, fixed zerofill, precision=2, dec sep align, clear infinite},
    columns/ask/.style={column name=$q^a$, fixed, fixed zerofill, precision=2, dec sep align, clear infinite},
    columns/Fplain/.style={column name=$F$, fixed, fixed zerofill, precision=2, dec sep align, clear infinite},
    columns/Fstag/.style={column name=$F^{\rm K}$, fixed, fixed zerofill, precision=2, dec sep align, clear infinite},
    columns/Fbenth/.style={column name=$F^{\rm B}$, fixed, fixed zerofill, precision=2, dec sep align, clear infinite},
    every head row/.style={before row=\toprule,after row=\midrule},
    every last row/.style={after row=\bottomrule},
]{\soybeanmkt}
\end{center}
\caption{Market data for Soybean observed on CBOE exchange on 15 December 2021. One-month contracts. The columns report from left to right contract identifier, bid and ask prices, model implied prices with and without seasonality management, classical Kriging implied prices.}
\label{tab:soybean}
\end{table}

\begin{figure}
\begin{center}
\scalebox{0.9}{%
\begin{tikzpicture}
\begin{axis}[xlabel=Maturity,
                    ylabel=Futures Price,
                    ylabel style={overlay},
                    xmin=-0.1, xmax=4.1,
                    xtick={0.085,1.085,2.085,3.085,4.088},
                    xticklabels={{\tt JAN22},{\tt JAN23},{\tt JAN24},{\tt JAN25},{\tt JAN26}},
                    ymin=1003, ymax=1297,
                    grid=major,
                    legend style={legend pos=south west},
                    axis background/.style={fill=gray!10}]
\addplot [color=mydarkblue,thick,smooth] table [y=Fstag1,x=T] from \soybeanmsh;
\addplot [color=mydarkblue,thick,smooth,style=dotted] table [y=Fplain1,x=T] from \soybeanmsh;
\addplot [color=mydarkmagenta,thick,smooth] table [y=Fbenth1,x=T] from \soybeanmsh;
\addplot [color=mydarkred,thick,only marks,mark=triangle,mark options={yshift=(\pgfplotmarksize+\pgflinewidth)/2}] table [y=Mbid1,x=T] from \soybeanmsh;
\addplot [color=mydarkred,thick,only marks,mark=triangle,mark options={yshift=(\pgflinewidth-\pgfplotmarksize)/2,style={rotate=180}}] table [y=Mask1,x=T] from \soybeanmsh;
\end{axis}
\end{tikzpicture}}
\hspace*{-0.25cm}
\scalebox{0.9}{%
\begin{tikzpicture}
\begin{axis}[xlabel=Maturity,
                    ylabel=Futures Price,
                    ylabel style={overlay},
                    xmin=-0.1, xmax=4.1,
                    xtick={0.085,1.085,2.085,3.085,4.088},
                    xticklabels={{\tt JAN22},{\tt JAN23},{\tt JAN24},{\tt JAN25},{\tt JAN26}},
                    ymin=1003, ymax=1297,
                    grid=major,
                    legend style={legend pos=south west},
                    axis background/.style={fill=gray!10}]
\addplot [color=mydarkblue,thick,smooth] table [y=Fstag3,x=T] from \soybeanmsh;
\addplot [color=mydarkblue,thick,smooth,style=dotted] table [y=Fplain3,x=T] from \soybeanmsh;
\addplot [color=mydarkmagenta,thick,smooth] table [y=Fbenth3,x=T] from \soybeanmsh;
\addplot [color=mydarkred,thick,only marks,mark=triangle,mark options={yshift=(\pgfplotmarksize+\pgflinewidth)/2}] table [y=Mbid3,x=T] from \soybeanmsh;
\addplot [color=mydarkred,thick,only marks,mark=triangle,mark options={yshift=(\pgflinewidth-\pgfplotmarksize)/2,style={rotate=180}}] table [y=Mask3,x=T] from \soybeanmsh;
\end{axis}
\end{tikzpicture}}
\\\vspace*{0.5cm}
\scalebox{0.9}{%
\begin{tikzpicture}
\begin{axis}[xlabel=Maturity,
                    ylabel=Futures Price,
                    ylabel style={overlay},
                    xmin=-0.1, xmax=4.1,
                    xtick={0.085,1.085,2.085,3.085,4.088},
                    xticklabels={{\tt JAN22},{\tt JAN23},{\tt JAN24},{\tt JAN25},{\tt JAN26}},
                    ymin=1003, ymax=1297,
                    grid=major,
                    legend style={legend pos=south west},
                    axis background/.style={fill=gray!10}]
\addplot [color=mydarkblue,thick,smooth] table [y=Fstag6,x=T] from \soybeanmsh;
\addplot [color=mydarkblue,thick,smooth,style=dotted] table [y=Fplain6,x=T] from \soybeanmsh;
\addplot [color=mydarkmagenta,thick,smooth] table [y=Fbenth6,x=T] from \soybeanmsh;
\addplot [color=mydarkred,thick,only marks,mark=triangle,mark options={yshift=(\pgfplotmarksize+\pgflinewidth)/2}] table [y=Mbid6,x=T] from \soybeanmsh;
\addplot [color=mydarkred,thick,only marks,mark=triangle,mark options={yshift=(\pgflinewidth-\pgfplotmarksize)/2,style={rotate=180}}] table [y=Mask6,x=T] from \soybeanmsh;
\end{axis}
\end{tikzpicture}}
\hspace*{-0.25cm}
\scalebox{0.9}{%
\begin{tikzpicture}
\begin{axis}[xlabel=Maturity,
                    ylabel=Futures Price,
                    ylabel style={overlay},
                    xmin=-0.1, xmax=4.1,
                    xtick={0.085,1.085,2.085,3.085,4.088},
                    xticklabels={{\tt JAN22},{\tt JAN23},{\tt JAN24},{\tt JAN25},{\tt JAN26}},
                    ymin=1003, ymax=1297,
                    grid=major,
                    legend style={legend pos=south west},
                    axis background/.style={fill=gray!10}]
\addplot [color=mydarkblue,thick,smooth] table [y=Fstag12,x=T] from \soybeanmsh;
\addplot [color=mydarkblue,thick,smooth,style=dotted] table [y=Fplain12,x=T] from \soybeanmsh;
\addplot [color=mydarkmagenta,thick,smooth] table [y=Fbenth12,x=T] from \soybeanmsh;
\addplot [color=mydarkred,thick,only marks,mark=triangle,mark options={yshift=(\pgfplotmarksize+\pgflinewidth)/2}] table [y=Mbid12,x=T] from \soybeanmsh;
\addplot [color=mydarkred,thick,only marks,mark=triangle,mark options={yshift=(\pgflinewidth-\pgfplotmarksize)/2,style={rotate=180}}] table [y=Mask12,x=T] from \soybeanmsh;
\end{axis}
\end{tikzpicture}}
\end{center}
\caption{Term structures of futures prices for Soybean quoted on 15 December 2021 on CBOE market. From top-left to bottom-right the panels show the delivery periods of one, three, six months and one year. Triangles represent bid-ask futures contracts prices. Continuous blue lines is our model with penalization, dotted blue lines without it, while continuous pink lines is classical Kriging.}
\label{fig:soybean}
\end{figure}

\begin{table}
\begin{center}
\pgfplotstabletypeset[
    col sep=tab,
    columns={id,bid,ask,Fplain,Fstag,Fbenth},
    columns/id/.style={column name=Contract, column type={c|}, string type},
    columns/bid/.style={column name=$q^b$, fixed, fixed zerofill, precision=2, dec sep align, clear infinite},
    columns/ask/.style={column name=$q^a$, fixed, fixed zerofill, precision=2, dec sep align, clear infinite},
    columns/Fplain/.style={column name=$F$, fixed, fixed zerofill, precision=2, dec sep align, clear infinite},
    columns/Fstag/.style={column name=$F^{\rm K}$, fixed, fixed zerofill, precision=2, dec sep align, clear infinite},
    columns/Fbenth/.style={column name=$F^{\rm B}$, fixed, fixed zerofill, precision=2, dec sep align, clear infinite},
    every head row/.style={before row=\toprule,after row=\midrule},
    every last row/.style={after row=\bottomrule},
]{\wheatmkt}
\end{center}
\caption{Market data for Wheat observed on CBOE exchange on 15 December 2021. One-month contracts. The columns report from left to right contract identifier, bid and ask prices, model implied prices with and without seasonality management, classical Kriging implied prices.}
\label{tab:wheat}
\end{table}

\begin{figure}
\begin{center}
\scalebox{0.9}{%
\begin{tikzpicture}
\begin{axis}[xlabel=Maturity,
                    ylabel=Futures Price,
                    ylabel style={overlay},
                    xmin=-0.1, xmax=3.1,
                    xtick={0.085,1.085,2.085,3.085},
                    xticklabels={{\tt JAN22},{\tt JAN23},{\tt JAN24},{\tt JAN25}},
                    ymin=653, ymax=807,
                    grid=major,
                    legend style={legend pos=south west},
                    axis background/.style={fill=gray!10}]
\addplot [color=mydarkblue,thick,smooth] table [y=Fstag1,x=T] from \wheatmsh;
\addplot [color=mydarkblue,thick,smooth,style=dotted] table [y=Fplain1,x=T] from \wheatmsh;
\addplot [color=mydarkmagenta,thick,smooth] table [y=Fbenth1,x=T] from \wheatmsh;
\addplot [color=mydarkred,thick,only marks,mark=triangle,mark options={yshift=(\pgfplotmarksize+\pgflinewidth)/2}] table [y=Mbid1,x=T] from \wheatmsh;
\addplot [color=mydarkred,thick,only marks,mark=triangle,mark options={yshift=(\pgflinewidth-\pgfplotmarksize)/2,style={rotate=180}}] table [y=Mask1,x=T] from \wheatmsh;
\end{axis}
\end{tikzpicture}}
\hspace*{0.25cm}
\scalebox{0.9}{%
\begin{tikzpicture}
\begin{axis}[xlabel=Maturity,
                    ylabel=Futures Price,
                    ylabel style={overlay},
                    xmin=-0.1, xmax=3.1,
                    xtick={0.085,1.085,2.085,3.085},
                    xticklabels={{\tt JAN22},{\tt JAN23},{\tt JAN24},{\tt JAN25}},
                    ymin=653, ymax=807,
                    grid=major,
                    legend style={legend pos=south west},
                    axis background/.style={fill=gray!10}]
\addplot [color=mydarkblue,thick,smooth] table [y=Fstag3,x=T] from \wheatmsh;
\addplot [color=mydarkblue,thick,smooth,style=dotted] table [y=Fplain3,x=T] from \wheatmsh;
\addplot [color=mydarkmagenta,thick,smooth] table [y=Fbenth3,x=T] from \wheatmsh;
\addplot [color=mydarkred,thick,only marks,mark=triangle,mark options={yshift=(\pgfplotmarksize+\pgflinewidth)/2}] table [y=Mbid3,x=T] from \wheatmsh;
\addplot [color=mydarkred,thick,only marks,mark=triangle,mark options={yshift=(\pgflinewidth-\pgfplotmarksize)/2,style={rotate=180}}] table [y=Mask3,x=T] from \wheatmsh;
\end{axis}
\end{tikzpicture}}
\\\vspace*{0.5cm}
\scalebox{0.9}{%
\begin{tikzpicture}
\begin{axis}[xlabel=Maturity,
                    ylabel=Futures Price,
                    ylabel style={overlay},
                    xmin=-0.1, xmax=3.1,
                    xtick={0.085,1.085,2.085,3.085},
                    xticklabels={{\tt JAN22},{\tt JAN23},{\tt JAN24},{\tt JAN25}},
                    ymin=653, ymax=807,
                    grid=major,
                    legend style={legend pos=south west},
                    axis background/.style={fill=gray!10}]
\addplot [color=mydarkblue,thick,smooth] table [y=Fstag6,x=T] from \wheatmsh;
\addplot [color=mydarkblue,thick,smooth,style=dotted] table [y=Fplain6,x=T] from \wheatmsh;
\addplot [color=mydarkmagenta,thick,smooth] table [y=Fbenth6,x=T] from \wheatmsh;
\addplot [color=mydarkred,thick,only marks,mark=triangle,mark options={yshift=(\pgfplotmarksize+\pgflinewidth)/2}] table [y=Mbid6,x=T] from \wheatmsh;
\addplot [color=mydarkred,thick,only marks,mark=triangle,mark options={yshift=(\pgflinewidth-\pgfplotmarksize)/2,style={rotate=180}}] table [y=Mask6,x=T] from \wheatmsh;
\end{axis}
\end{tikzpicture}}
\hspace*{0.25cm}
\scalebox{0.9}{%
\begin{tikzpicture}
\begin{axis}[xlabel=Maturity,
                    ylabel=Futures Price,
                    ylabel style={overlay},
                    xmin=-0.1, xmax=3.1,
                    xtick={0.085,1.085,2.085,3.085},
                    xticklabels={{\tt JAN22},{\tt JAN23},{\tt JAN24},{\tt JAN25}},
                    ymin=653, ymax=807,
                    grid=major,
                    legend style={legend pos=south west},
                    axis background/.style={fill=gray!10}]
\addplot [color=mydarkblue,thick,smooth] table [y=Fstag12,x=T] from \wheatmsh;
\addplot [color=mydarkblue,thick,smooth,style=dotted] table [y=Fplain12,x=T] from \wheatmsh;
\addplot [color=mydarkmagenta,thick,smooth] table [y=Fbenth12,x=T] from \wheatmsh;
\addplot [color=mydarkred,thick,only marks,mark=triangle,mark options={yshift=(\pgfplotmarksize+\pgflinewidth)/2}] table [y=Mbid12,x=T] from \wheatmsh;
\addplot [color=mydarkred,thick,only marks,mark=triangle,mark options={yshift=(\pgflinewidth-\pgfplotmarksize)/2,style={rotate=180}}] table [y=Mask12,x=T] from \wheatmsh;
\end{axis}
\end{tikzpicture}}
\end{center}
\caption{Term structures of futures prices for Wheat quoted on 15 December 2021 on CBOE market. From top-left to bottom-right the panels show the delivery periods of one, three, six months and one year. Triangles represent bid-ask futures contracts prices. Continuous blue lines is our model with penalization, dotted blue lines without it, while continuous pink lines is classical Kriging.}
\label{fig:wheat}
\end{figure}

\section{Conclusion and further developments}

In this paper we investigated a Bayesian technique known as Kriging to build the futures term structure of commodity assets. We were able to include in our analysis seasonalities and to consider bid-ask spreads of market quotations. We discussed many examples in various commodity markets on different observation dates. Further developments of our research concern applying this technique to other asset classes such as inflation which usually present a complex form for futures term structures. Moreover, we plan to investigate the seasonality treatment proposed for futures prices also in the case of bootstrapping their volatility surface.

\printbibliography

\end{document}